%% file: bridge_correlation_ACT.tex
\documentclass[fleqn,usenatbib]{mnras}

\usepackage{newtxtext,newtxmath}

\usepackage[T1]{fontenc}

\DeclareRobustCommand{\VAN}[3]{#2}
\let\VANthebibliography\thebibliography
\def\thebibliography{\DeclareRobustCommand{\VAN}[3]{##3}\VANthebibliography}

\usepackage{graphicx}	
\usepackage{amsmath}	
\usepackage{amsmath}
\usepackage{float}
\usepackage{booktabs}
\usepackage{textcomp} 
\usepackage{gensymb}
\usepackage[utf8]{inputenc}
\usepackage{multirow}
\usepackage{rotating}
\usepackage{savesym}
\savesymbol{tablenum} 
\usepackage{siunitx}
\restoresymbol{SIX}{tablenum}
\usepackage{tabularx}
\usepackage{tikz}
\usepackage{ulem}
\usepackage{xifthen}
\usepackage{xspace}
\usepackage{soul}

\newcommand{\Planck}{\textit{Planck}\xspace}

\newcommand{\XMM}{\textit{XMM-Newton}\xspace}

\DeclareSIUnit{\parsec}{pc}
\DeclareSIUnit{\arcminute}{arcmin}
\DeclareSIUnit{\year}{yr}

\newenvironment{enumerate*}{\begin{enumerate}%
        \setlength{\itemsep}{0pt}%
        \setlength{\parskip}{0pt}%
    }{\end{enumerate}}

\newcommand{\verify}[1]{{\color{Sepia} #1}}

\newcommand{\invisible}[1]{}

\newcommand{\const}[3]{\newcommand{#1}[1][]{{\verify{\ensuremath{#2\ifthenelse{\isempty{##1}}{\,#3}{}}}}}}

\title[Thermal and non-thermal components in A399--A401]{The thermal and non-thermal components within and between galaxy clusters Abell 399 and Abell 401}

\input{authors}

\date{Accepted XXX. Received YYY; in original form ZZZ}

\pubyear{2022}

\begin{document}
\label{firstpage}
\pagerange{\pageref{firstpage}--\pageref{lastpage}}
\maketitle

\begin{abstract}
    We measure the local correlation between radio emission and Compton-$y$ signal across two galaxy clusters, Abell~399 and Abell~401, using maps from the Low-Frequency Array (LOFAR) and the Atacama Cosmology Telescope (ACT) + \Planck. These datasets allow us to make the first measurement of this kind at $\sim$arcminute resolution. We find that the radio brightness scales as $F_{\mathrm{radio}} \propto y^{1.5}$ for Abell~401 and $F_{\mathrm{radio}} \propto y^{2.8}$ for Abell~399. Furthermore, using \XMM data, we derive a sublinear correlation between radio and X-ray brightness for both the clusters ($F_{\mathrm{radio}} \propto F_{\rm X}^{0.7}$). Finally, we correlate the Compton-$y$ and X-ray data, finding that an isothermal model is consistent with the cluster profiles, $y \propto F_{\rm X}^{0.5}$. By adopting an isothermal--$\beta$ model, we are able, for the first time, to jointly use radio, X-ray, and Compton-$y$ data to estimate the scaling index for the magnetic field profile, $B(r) \propto n_{\mathrm{e}}(r)^{\eta}$ in the injection and re-acceleration scenarios. Applying this model, we find that the combined radio and Compton-$y$ signal exhibits a significantly tighter correlation with the X-ray across the clusters than when the datasets are independently correlated. We find $\eta \sim 0.6{-}0.8$. These results are consistent with the upper limit we derive for the scaling index of the magnetic field using rotation measure values for two radio galaxies in Abell~401. We also measure the radio, Compton-$y$, and X-ray correlations in the filament between the clusters but conclude that deeper data are required for a convincing determination of the correlations in the filament.
    \end{abstract}




\section{Introduction}

Galaxy clusters are giant laboratories for plasma physics that allow us to study how thermal and non-thermal ({\it e.g.}, magnetic fields and relativistic electrons) components are related on scales of $\sim$Mpc. A growing number of clusters show the presence of diffuse radio emission on Mpc scales with low surface brightness emission at the cluster center \citep[e.g.,][and references therein]{Murgia09, vWeeren19}.
These extended sources are also known as radio halos and are typically found in clusters that show ongoing merger activity \citep[e.g.,][and references therein]{Cuciti21}.

A correlation between the integrated radio power over the entire cluster volume at 1.4 GHz, due to synchrotron emission, and X-ray luminosity, due to thermal bremsstrahlung emission, is known to exist in galaxy clusters hosting radio halos \citep[see e.g.,][]{giovannini00, feretti12}. A similar correlation has been discovered by \cite{basu12} and confirmed by \cite{Cassano13} and \cite{sommer_Basu14} between radio halo power at 1.4 GHz and the integrated Compton-$y$ signal due to the Sunyaev-Zeldovich effect \citep[][SZ]{sz72} produced by inverse Compton scattering of the cosmic microwave background (CMB) photons with thermal free electrons in the intracluster medium (ICM). The thermal SZ effect is proportional to the Compton-$y$ parameter, defined as \citep{Carlstrom2002}:
\begin{equation}
    y =  \frac{\sigma_\textsc{t}}{m_{\rm e}c^2} \int n_{\rm e} (l) k T(l) dl,
    \label{eq:tsz}
\end{equation}
where $\sigma_\textsc{t}$ is the Thomson cross section, $m_{\rm e}$ is the electron mass, $c$ the speed of the light, $k$ the Boltzmann constant, $l$ is the distance along the line of sight, $T(l)$ is the electron temperature and $n_{\mathrm{e}}(l)$ is the electron density. An analogous correlation has been discovered by \cite{weeren21} using 150 MHz data from Low-Frequency Array (LOFAR) and \Planck PSZ2 clusters \citep{Planck2016}. Since synchrotron emission depends on the non-thermal particle pressure and magnetic field energy density, and the bremsstrahlung emission is proportional to the square of the gas density, the radio--X-ray (radio--X hereafter) relation links non-thermal pressure and gas mass. Both thermal and non-thermal electrons can scatter the CMB photons, see \cite{Birkinshaw1999} for further details. However the SZ effect arising from the thermal electron population is dominant mainly due to the lower density of relativistic than non-relativistic electrons. We conclude that the radio--SZ correlation links non-thermal components to thermal electron pressure.

Previous work has studied the integrated signals of galaxy clusters hosting radio halos, and the radio--X correlation has also been found locally in some galaxy clusters containing radio halos in a point-by-point correlation \citep{Govoni01}. Most of the studies available in the literature on the radio--SZ correlation refer only to integrated cluster properties. The galaxy clusters RXJ1347.5-1145 and Coma are the only cases where a local radio--SZ connection has been investigated and found. In the first cluster, \cite{Ferrari11} examined a currently unique correspondence, noted earlier in \cite{Mroczkowski2011}, between the excess emission in the radio mini-halo at the center of RXJ1347.5-1145 and a high-pressure region revealed by SZ observations obtained with MUSTANG \citep{Mason10}. Taking advantage of the large angular extent of Coma, \cite{Planck13_Coma} found a point-by-point spatial correlation between the new Planck SZ map and ancillary radio maps at $10^{\prime}$ angular resolution in the Coma cluster.

The origin of the relativistic electrons responsible for the radio halo emission is not completely understood and is still a matter of debate. Two main theories have been formulated and studied in the last decades. The first one is based on the idea that pre-existing electrons, injected by starburst activity, violent galactic winds, and active galactic nuclei, are systematically re-accelerated in the ICM due to turbulence produced by merger or post-merger shocks \citep[e.g.,][]{brunetti01, gitti02}. This model is known as the re-acceleration scenario because the number of relativistic electrons remains constant. Because merger events produce re-acceleration of relativistic electrons, this model can naturally explain the observed connection between radio halos and merging clusters. The second hypothesis is based on the idea that non-thermal electrons are continuously injected in the ICM \citep{dennison80}; it is known as an injection scenario. This theory solves the observed large spatial size of the radio halos in a simple way: relativistic electrons are produced over the radio halo volume by collisions between long-lived relativistic protons and thermal ions in the ICM. The resulting synchrotron radiation is emitted near the region where new electrons are injected. However, the injection model does not easily explain the correlation between cluster mergers and radio halos. Although the two models are very different, hybrid models have been proposed where the non-thermal electron population is a consequence of the combination of the previous two models. See \cite{Zandanel2014} for more information regarding hybrid models.

In this work, we study the radio--X, SZ--X, and, for the first time with an ${\sim}$arcminute angular resolution, the radio--SZ correlation locally in a pair of galaxy clusters and in the filament between these clusters. We use an Atacama Cosmology Telescope (ACT)+\textit{Planck} map for the SZ \citep{hincks22}, a 140 MHz LOFAR map for the radio \citep{Govoni19}, and four \XMM pointings in the energy band between 0.4 and 7.2 keV for the X-ray counterpart. Among the main advantages of using higher angular resolution maps, we cite the greater number of independent points to be correlated and the identification and removal of compact background sources in the field. The perfect laboratory for our purpose is the system comprising Abell 399 (A399) and Abell 401 (A401), with redshifts $z = $ 0.072--0.074 \citep{Oeg01}. The two clusters are separated on the plane of the sky by $36.9^{\prime}$, or, equivalently, by 3.2\,Mpc in the cosmology adopted in this paper.\footnote{We assume a flat $\Lambda$CDM Universe with $H_{0}=67.6$ km s$^{-1}$ Mpc$^{-1}$, $\Omega _{m} = 0.31$ and $\Omega _{\Lambda} = 0.69$ \citep{aiola20}.} \cite{sake04} derived the temperature of the two clusters: 7.2\,keV for A399 and 8.5\,keV for A401.
Recently, \cite{hincks22} concluded that most of the separation of the two clusters is along the line of sight, and estimated that their physical separation is actually $11.1^{+3.2} _{-2.6}$\,Mpc. Both A399 and A401 host radio halos \citep{murgia2010} and several observations at different wavelengths have detected signals from the low density and high temperature gas that resides between the two clusters ({\it e.g.}, \citealt{sake04, Planck13, akam17, Bonjean2018, Govoni19, hincks22}, and references therein). The latter authors estimated an average gas density in the filament of $\left( 0.88 \pm 0.24 \right) \times 10^{-4}$ cm$^{-3}$; given the low gas density in the filament, the radio emission we observe in this system requires second-order Fermi re-acceleration mechanisms, as first suggested by \cite{brunetti_vazza2020}.
In the literature, there is widespread agreement that the two clusters are in a pre-merger phase \citep{Fujita1996, Markevitch1998, sake04}, in contrast to the earlier hypothesis by \cite{Fabian97} suggesting a post-merger state. The irregularities in the X-ray morphology in A399 and A401 can be explained with minor mergers in the two clusters \citep{Bourdin08}. This hypothesis, for A401, appears to be confirmed by the SZ analysis performed in \cite{hincks22} using Green Bank Telescope MUSTANG-2 observations.

In this paper, we provide an overview of the data we use in Sec.~\ref{Sect:data}. In Sec. \ref{Sect:Correlation} we describe the method used to extract the radio, X-ray and SZ data starting from the three maps, then we determine the radio--X, radio--SZ, and SZ--X correlations in the two clusters and the intercluster region. Sec. \ref{Sect:Interpretations} discusses previous results for the three correlations in the two clusters, mainly focusing on the radio--X and radio--SZ correlations. We show how the SZ-X correlation and the combination of the others two correlations can be used independently to derive the clusters' thermodynamic properties and probe the link between the thermal and non-thermal components in the ICM. In Sec. \ref{Sec:Magnetic_Field_Slope} we derive magnetic field properties in the two clusters. By jointly using radio, X-ray, and SZ data, we estimate the scaling index of the magnetic field in the two galaxy clusters for the injection and re-acceleration scenarios. We compare this result to an independent, albeit approximate, estimate based on ancillary rotation measure data available for A401. Finally we conclude in Section \ref{Sect:Conclusions}.

\section{DATA}
\label{Sect:data}

\begin{figure}
    \centering
    {\includegraphics[trim={0.6cm 1.55cm 0.45cm 0.6cm},clip,scale=0.83]{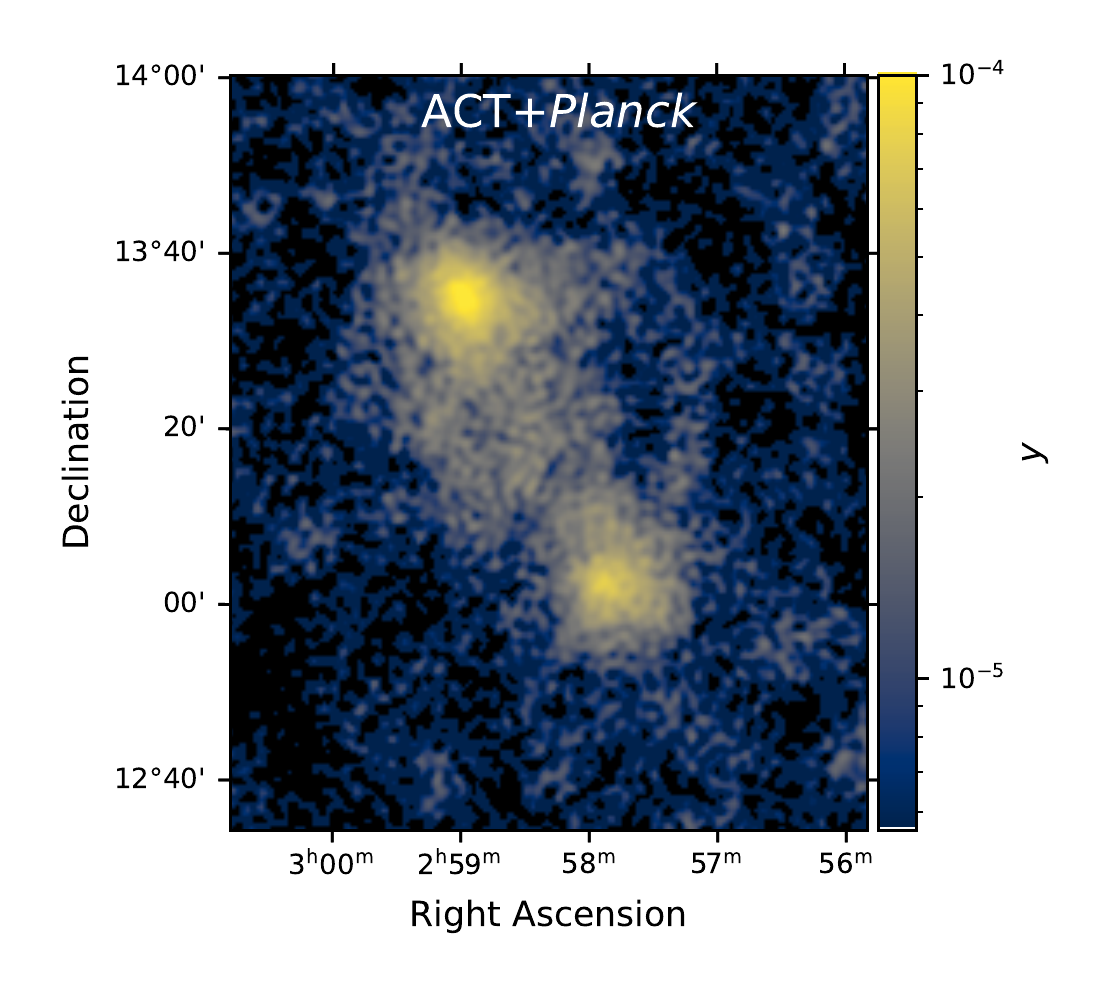}} 
    {\includegraphics[trim={0.6cm 1.55cm 0.45cm 0.6cm},clip,scale=0.83]{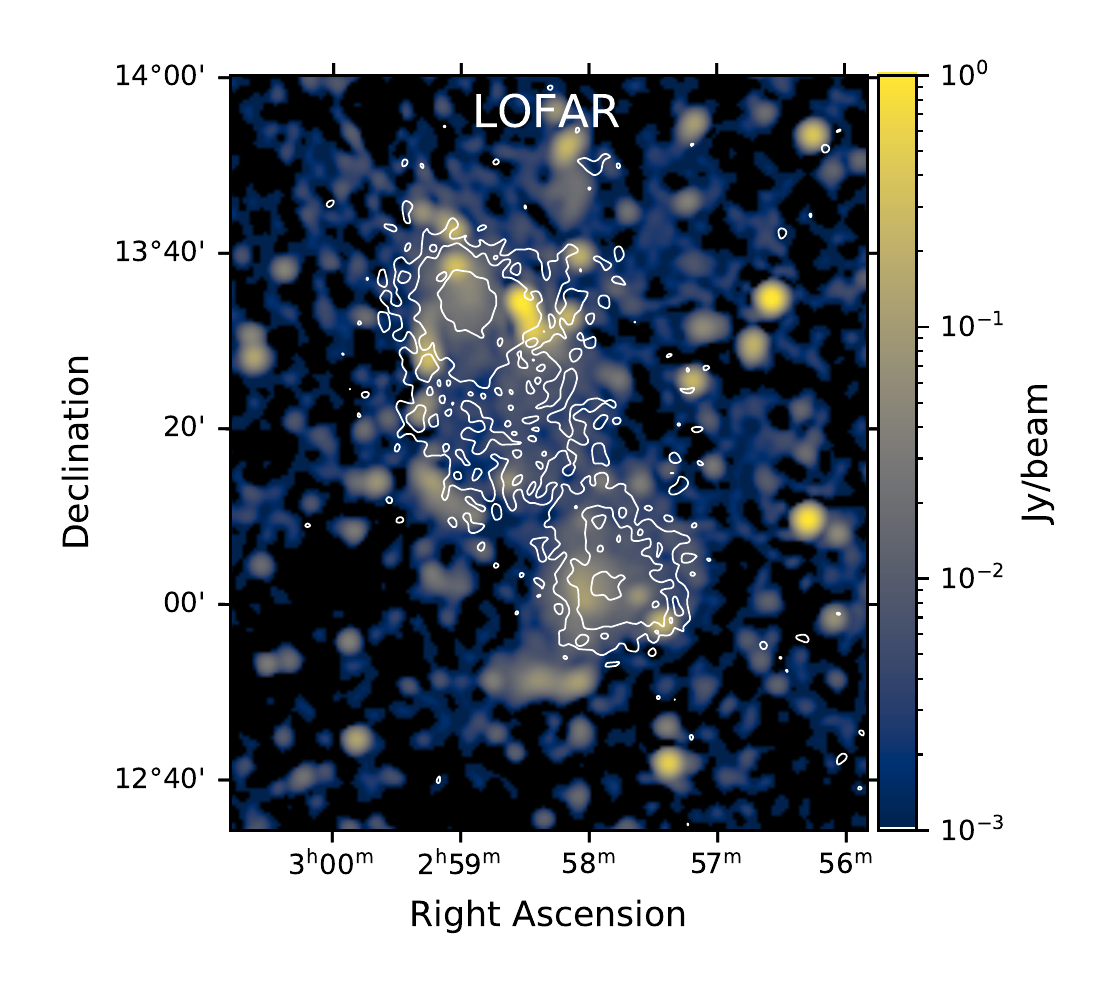}}
    {\includegraphics[trim={0.6cm 0.6cm 0.6cm 0.6cm},clip,scale=0.83]{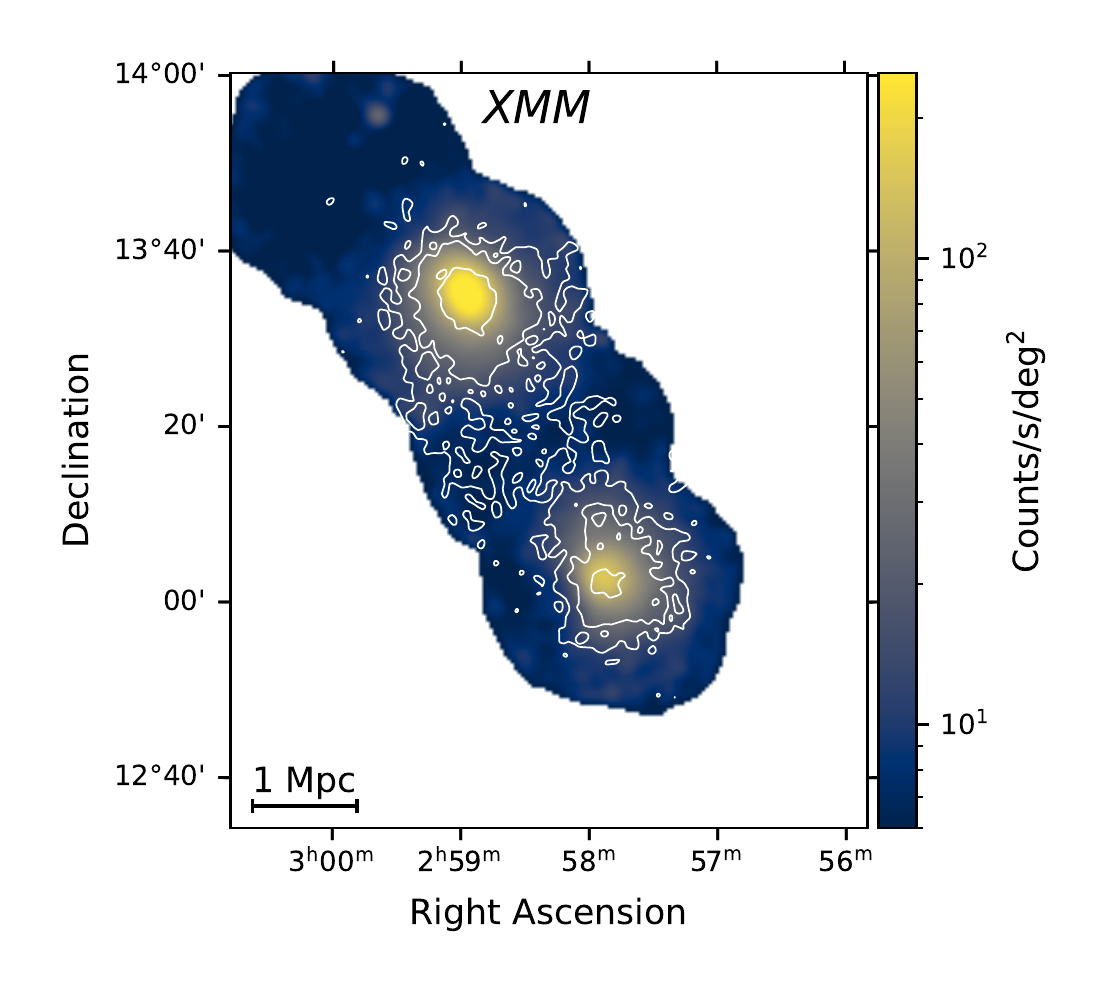}}
    \caption{{\bf Top}: ACT+\Planck Compton-$y$ map of the cluster pair A401--A399 (north and south, respectively) at $1.65^{\prime}$ resolution. 
    {\bf Middle:} 140 MHz LOFAR map convolved to $1.65^{\prime}$.
    {\bf Bottom}: \XMM map using four pointings in the 0.4--1.25+2.0--7.2 keV band. The \XMM map has been corrected for background and exposure, and smoothed to $1.65^{\prime}$. White contours superimposed on the LOFAR and \XMM maps refer to the 3, 5, and 10$\sigma$ ACT levels. The physical scale on the plane of the sky is shown on the bottom left side of the \XMM map.
    }
    \label{fig:Maps}
\end{figure}

\begin{figure*}
    {\includegraphics[scale=0.53]{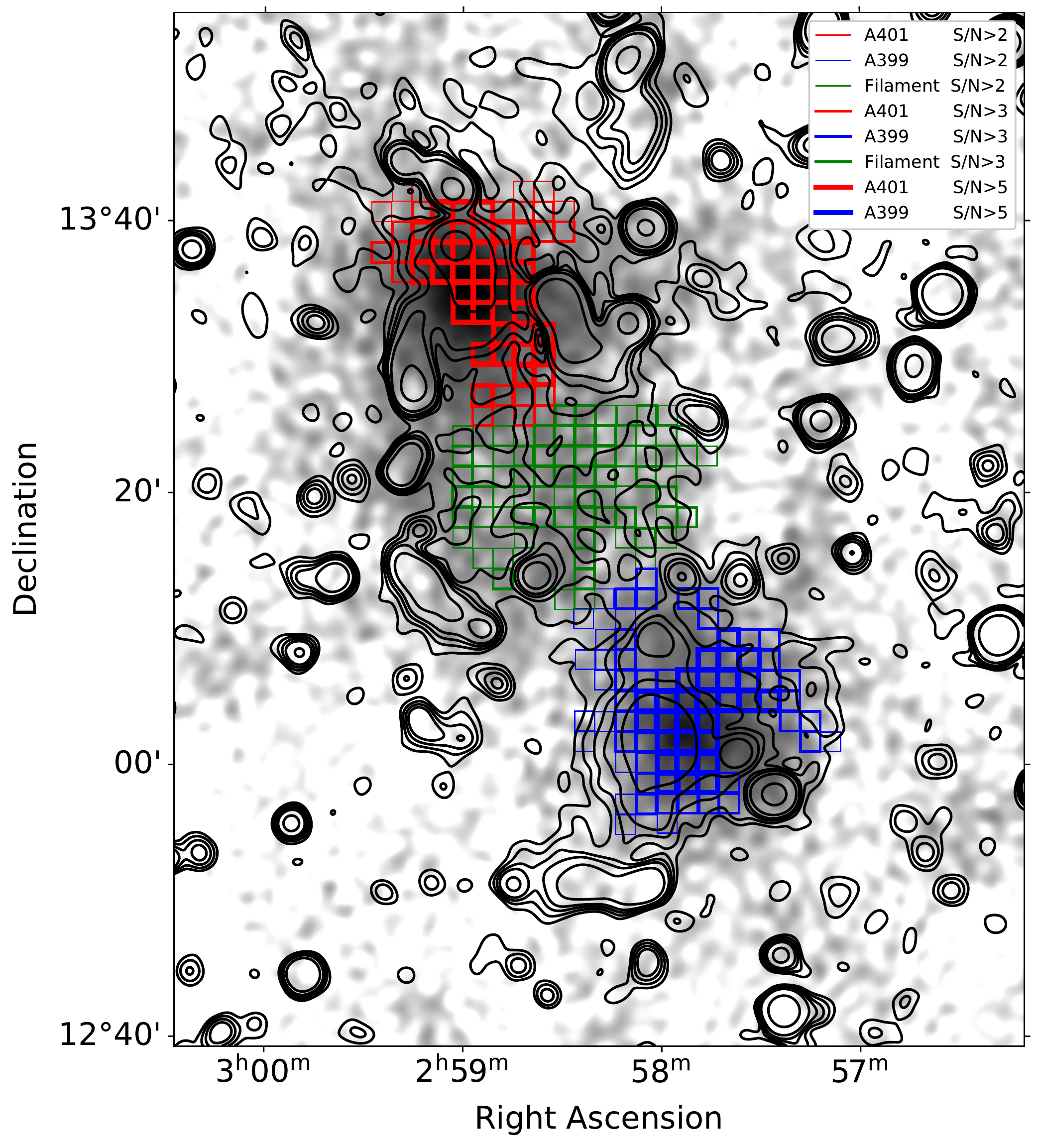}}
    \caption{ACT+\Planck Compton-$y$ map with the unmasked regions used in Sec. \ref{SubSec:SZ-radio} to study radio--SZ spatial correlations represented by colored boxes. 
    Data points were extracted from each map using the boxes shown here (each of size $1.65^{\prime}$ by $1.65^{\prime}$). This size corresponds to the effective resolution of the maps (see text). Red, blue, and green boxes refer respectively to A401, A399, and the filament region. As we show in the legend, boxes with thinner lines refer to the regions that satisfy weaker $S/N$ conditions. Note that each box corresponds to a single measurement in Fig. \ref{fig:Correlation_SZ_radio} and \ref{fig:Correlation_Bridge}. The grid excludes radio galaxies and candidate radio relics visible in the LOFAR map, traced by black contour levels, and regions with $S/N<2$ in one of the two maps. LOFAR contour levels are: 0.003, 0.006, 0.01, 0.015 and 0.03 Jy/beam. We note the offset along the east-west line between the radio brightness and the SZ peaks in A399.}
    \label{fig:Masks}
\end{figure*}

Here we present the data used to analyze spatial correlations in the two clusters and in the region between them. The highest angular resolution and depth maps available were used to highlight structures in the diffuse emission.

For the SZ data, we make use of the ACT+\textit{Planck} map of the A399--A401 system presented in \cite{hincks22}. ACT is a 6-meter off-axis Gregorian telescope observing since 2007 from the Atacama Desert at an altitude of 5190 meters \citep{Thornton16}. The map includes ACT data at 98, 150, and 220 GHz from 2008 up to 2019, together with \textit{Planck} data to improve sensitivity at scales larger than $ \gtrsim 6^{\prime} $, where ACT data, for the map we use in this paper, can be contaminated by the atmosphere; the \textit{Planck} 353 and 545 GHz channel maps are included to remove Galactic dust and cosmic infrared background emission. The main difference with respect to ACT Data Release 5 \citep{naess/etal:2020} is the inclusion of data acquired during 2019. The Compton-$y$ map was made using an internal linear combination (ILC) of the single-frequency maps, assuming the non-relativistic SZ formula, in a manner similar to \cite{madhavacheril/etal:2020}; see \cite{hincks22} for details. The sensitivity of the map in Compton-$y$ units is $5.9 \times 10^{-6} $ per effective beam ($1.65'$). The map is shown on the top panel of Fig. \ref{fig:Maps}.

For the radio counterpart we use the LOFAR 140\,MHz map at $80^{\prime \prime}$ angular resolution (full width half maximum) presented by \cite{Govoni19}. LOFAR is an interferometric array of radio telescopes located mainly in the Netherlands; since 2012 it has mapped the sky at frequencies between 10 and 240\,MHz \citep{Lofar2013}. The observations used here were carried out under project LC2\_005, using the full Dutch Array (24 core stations and 14 remote stations) in the high band antenna (HBA) dual inner configuration. Data were recorded with an integration time of 1\,s and with the 8-bit mode in order to have a continuous frequency coverage between 110 and 190\,MHz. With a total observing time of 10 hours with the HBA stations, the sensitivity was 1.0 mJy/beam, close to the expected confusion noise limit at this frequency and resolution. The LOFAR map, convolved to the ACT $1.65^{\prime}$ resolution using a Gaussian kernel, is shown on the middle panel in Fig. \ref{fig:Maps}.

The X-ray map is a mosaic of four \XMM pointings centred on A399, the intercluster region, A401, and a northeast field with respect to A401. The observation IDs of each pointing are 0112260101, 0112260201, 0112260301, and 0112260401 and are the same raw data used by \cite{sake04} to study the two clusters and the high-temperature gas between them. The two MOS receivers \citep{mos2001} and the PN detector \citep{pn2001} have an angular resolution of FWHM $\sim 6 ^{\prime \prime}$ at the center of the field of view. We analyzed the 0.4--1.25 and 2.0--7.2 keV energy bands in order to exclude the most contaminated energies (1.25--2.0 keV) by the instrumental lines and the residual contamination by soft proton flares at high energies. The final image is a combination of the two different bands. 
All the data sets were reduced using the \XMM Extended Source Analysis Software (ESAS) package \citep{Snowden04} for the analysis of imaging MOS and PN mode observations. For each observation, we followed the procedure described in the XMM ESAS Cookbook\footnote{https://heasarc.gsfc.nasa.gov/FTP/xmm/software/xmm-esas/xmm-esas-v13.pdf} to construct a map for each pointing. The background spectrum was estimated using the \textit{mos\_back} and \textit{pn\_back} tasks in regions where the contamination from the clusters or the filament is negligible. We estimated the soft proton contamination fitting the spectra of the empty regions with a broken power-law using the \textit{Xspec} software \citep{xspec96}. The point sources in the X-ray images have been excluded using the ESAS \textit{cheese} routine. The resulting point sources mask was modified to prevent blanking of the brightest regions of the two clusters. The mosaic of the four observations, background subtracted, corrected by the exposure time, without point sources, and then smoothed to the resolution of the Compton-$y$ ACT+\Planck map, is shown in the bottom panel of Fig.~\ref{fig:Maps}.

\section{CORRELATION STUDIES}
\label{Sect:Correlation}

In this section, we study the radio--SZ, radio--X, and SZ--X spatial correlations. For a proper comparison, we first smoothed the LOFAR 140\,MHz and the XMM maps to the $1.65^{\prime}$ ACT+\Planck map effective resolution. We then projected the ACT+\Planck and the XMM-Newton maps to LOFAR geometry (1.5$^{\prime \prime}$ pixel size) using \textsc{Montage}.\footnote{http://montage.ipac.caltech.edu/index.html} Note that the pixelization does not affect the results we will present in the following; using the ACT+\Planck geometry, we get compatible results. Fig. \ref{fig:Maps} shows the images used for our analysis. Measuring the level of the fluctuations in the ACT+\Planck and LOFAR maps, we find $5.9 \times 10^{-6}$ and 1.0 mJy/beam, respectively. For the \XMM map, we measure the X-ray background, corrected for the exposure time, in the north-east field with respect to A401, finding 4.2\,counts\,s$^{-1}$\,deg$^{-2}$.

We create a grid to study the correlations inside the two galaxy clusters and in the intercluster region. Since we are interested in the radio halos and the emission of the filament, we exclude from the grid the radio galaxies and the candidate radio relics detected in the LOFAR 140 MHz \citep{Govoni19} and 1.4 GHz Very Large Array \citep{murgia2010} images. Compact sources in the filament region detected by MUSTANG-2, which do not have a clear correspondence in the radio maps, have also been masked.
The resulting grid for each correlation is created after a signal-to-noise $S/N <2$ cut on both the images. We investigate how the threshold choice could impact the results by also making grids with $S/N$ cuts below 3 and 5. The results are consistent within $3 \sigma$ for all the adopted $S/N$ thresholds, except for the radio--SZ correlation of A399; see the discussion in Sec.~\ref{SubSec: radio-sz interp}.
The fits were done using \textsc{emcee} \citep{foreman13}, a Python implementation of Goodman \& Weare’s affine invariant Markov chain Monte Carlo (MCMC) ensemble sampler \citep{goodman10}.

The resulting grid for the radio--SZ correlation and for the different adopted $S/N$ thresholds is shown in Fig.~\ref{fig:Masks} and over-plotted on the ACT+\textit{Planck} map. We use boxes of size $1.65^{\prime}$ that corresponds to the effective resolution of the maps. The red, blue and green boxes cover A401, A399 and the bridge region, respectively. We consider the boundaries of the two clusters to be at four times the core radii estimated in \cite{hincks22}. This condition ensures that the best fit model contribution of the filament is negligible in the boxes that belong to the two clusters. We also extracted a grid where the clusters extend up to five times the core radii. By increasing the clusters' boundaries by 25\% ($\times 5 r_{\rm c}$), the results we derive are $1\sigma$ compatible and vary by 8\% at most with respect to the values reported in Table \ref{tab:fit_params}. We conclude that the choice of the boundaries of the clusters does not introduce any systematic in our analysis. However, we adopt the grid shown in Fig.~\ref{fig:Masks} to minimize the contribution of the filament in the outer boxes of the two clusters. Each box corresponds to a single point at which the correlation is calculated. Note that the grid shown in Fig. \ref{fig:Masks} refers only to the radio--SZ correlation. For the radio--X and SZ--X point-by-point comparison, the masks we used differ only in the intercluster region, where fewer points satisfy the threshold condition, mainly due to low X-ray brightness in the region.

\begin{figure}
    \centering
    {\includegraphics[scale=0.5]{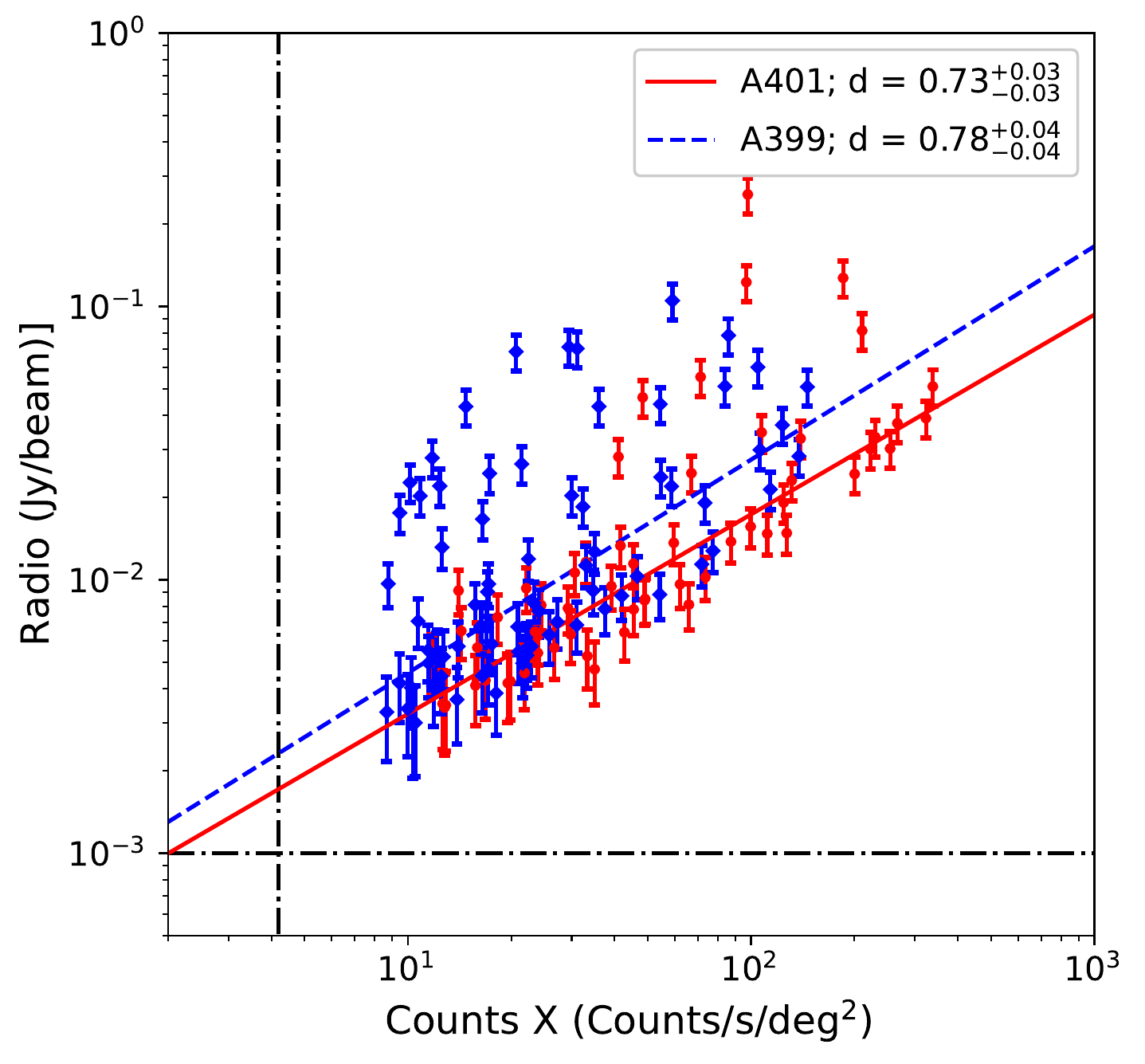}}
    {\includegraphics[scale=0.5]{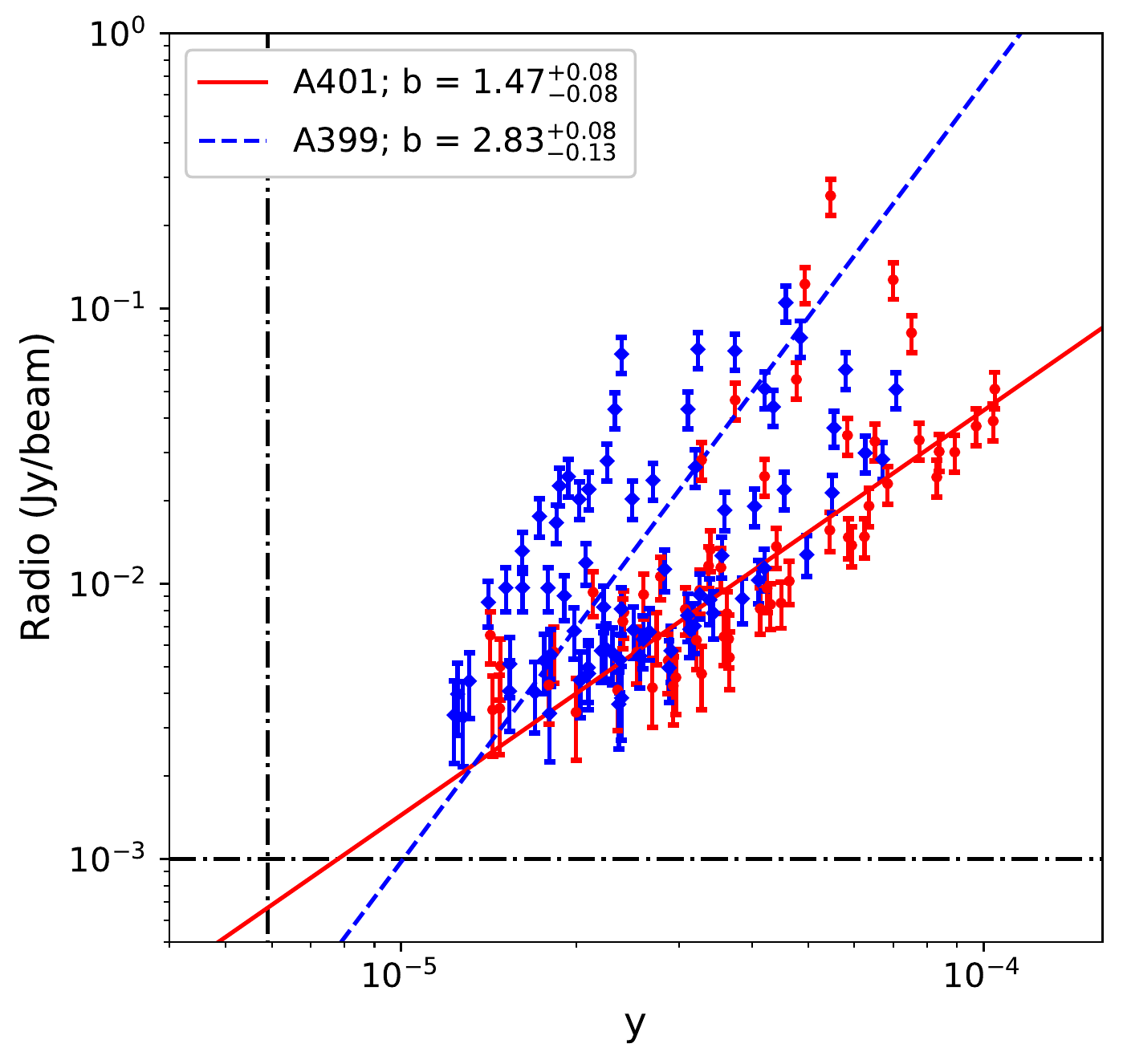}} 
    {\includegraphics[scale=0.5]{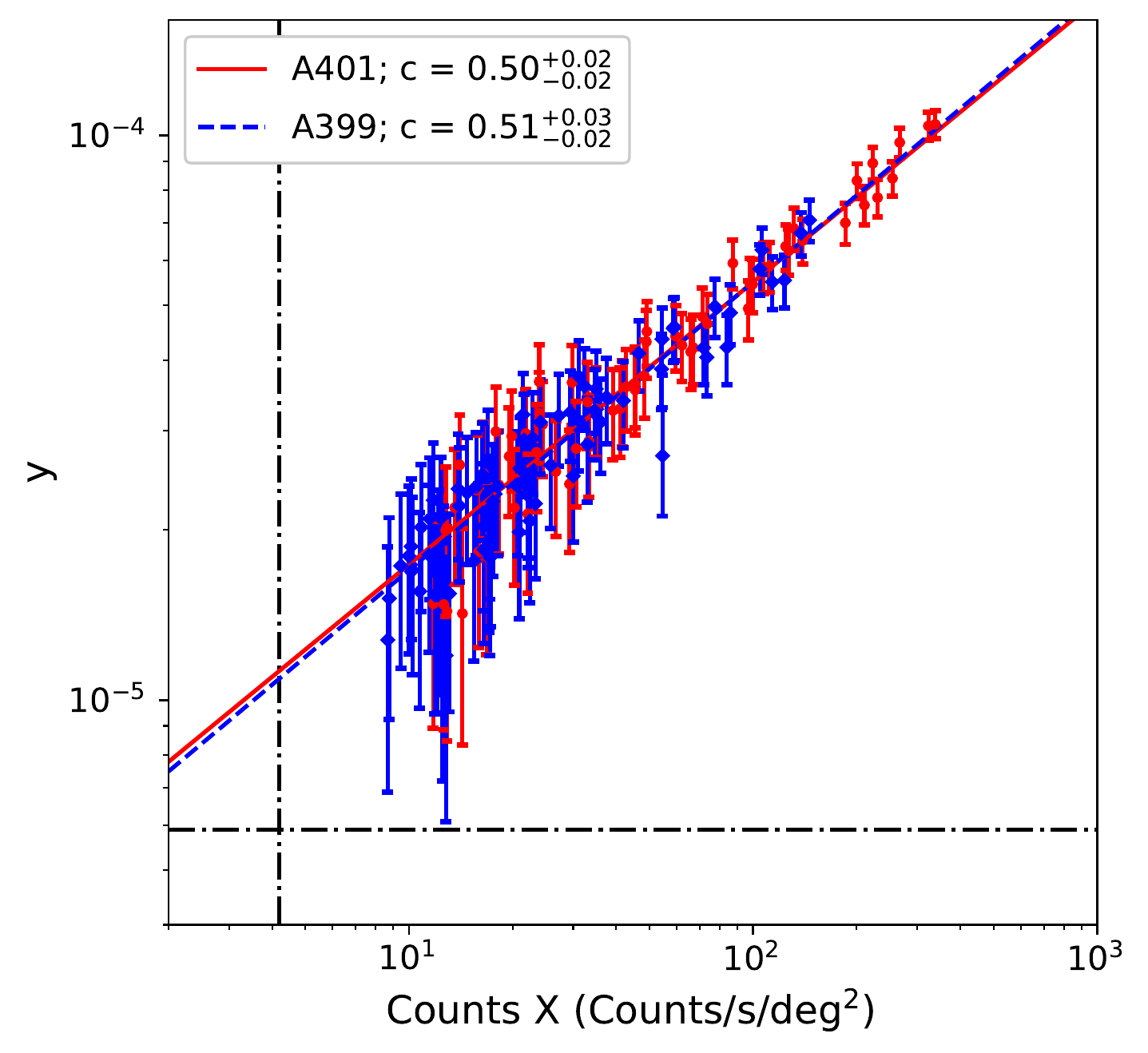}}
    \caption{\textbf{Top}: Correlation between the radio and X-ray count rate maps. \textbf{Middle}: Correlation between the radio map and the $y$ signal. \textbf{Bottom:} Correlation between the $y$ signal and the X-ray count rate map. The red points and the solid line refer to A401, while the blue diamond markers and dashed line refer to A399. In the three figures, we show only the points that satisfy the condition $S/N > 2$. The horizontal and vertical dash-dotted lines indicate the noise level in the two maps. Error bars in the x-axes are not included in the panels to make the figures easier to read but have been considered in estimating the best-fit parameters.}
    \label{fig:Correlation_SZ_radio}
\end{figure}

\begin{figure*}
    \centering
    {\includegraphics[scale=0.4]{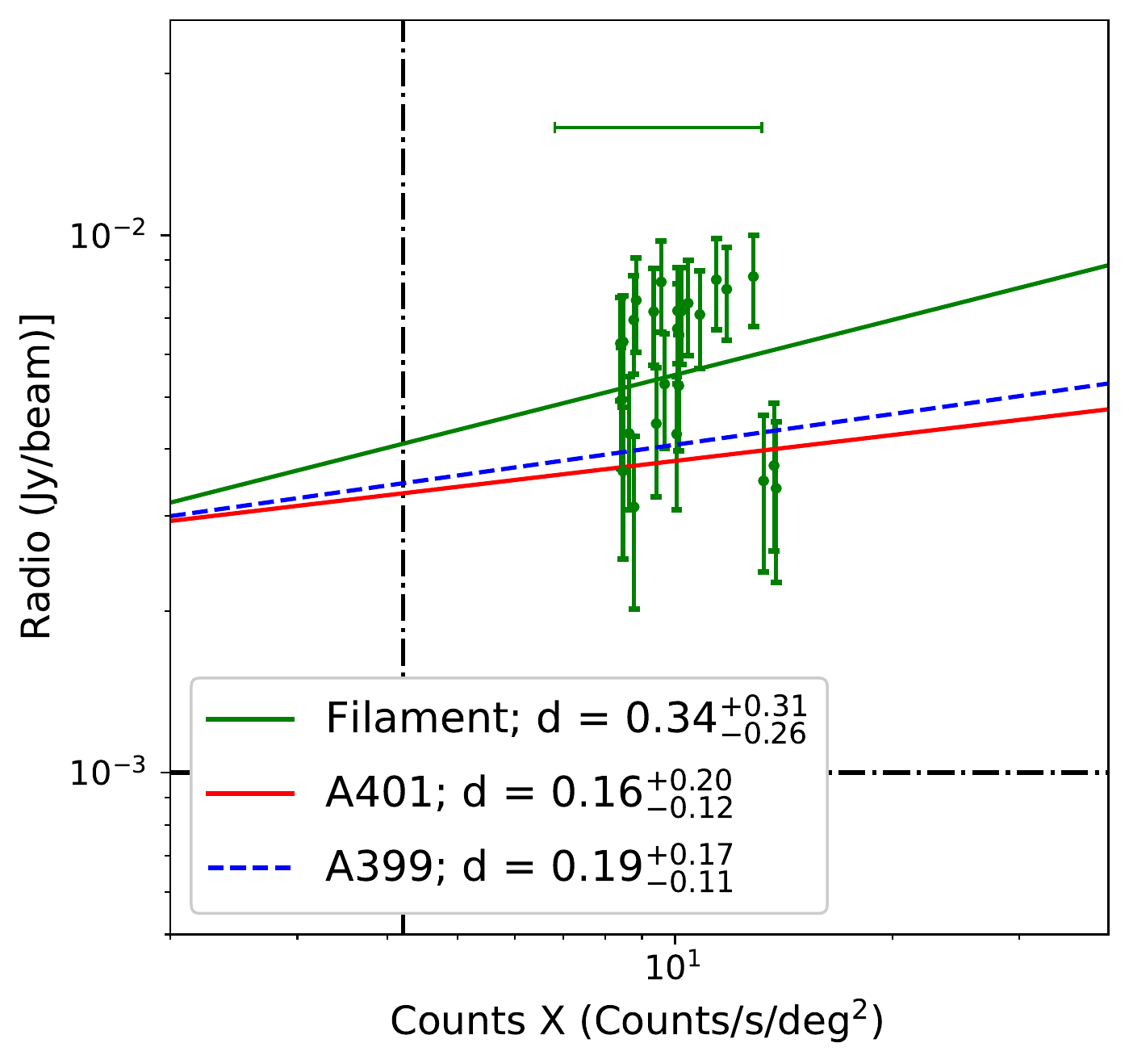}}
    {\includegraphics[scale=0.4]{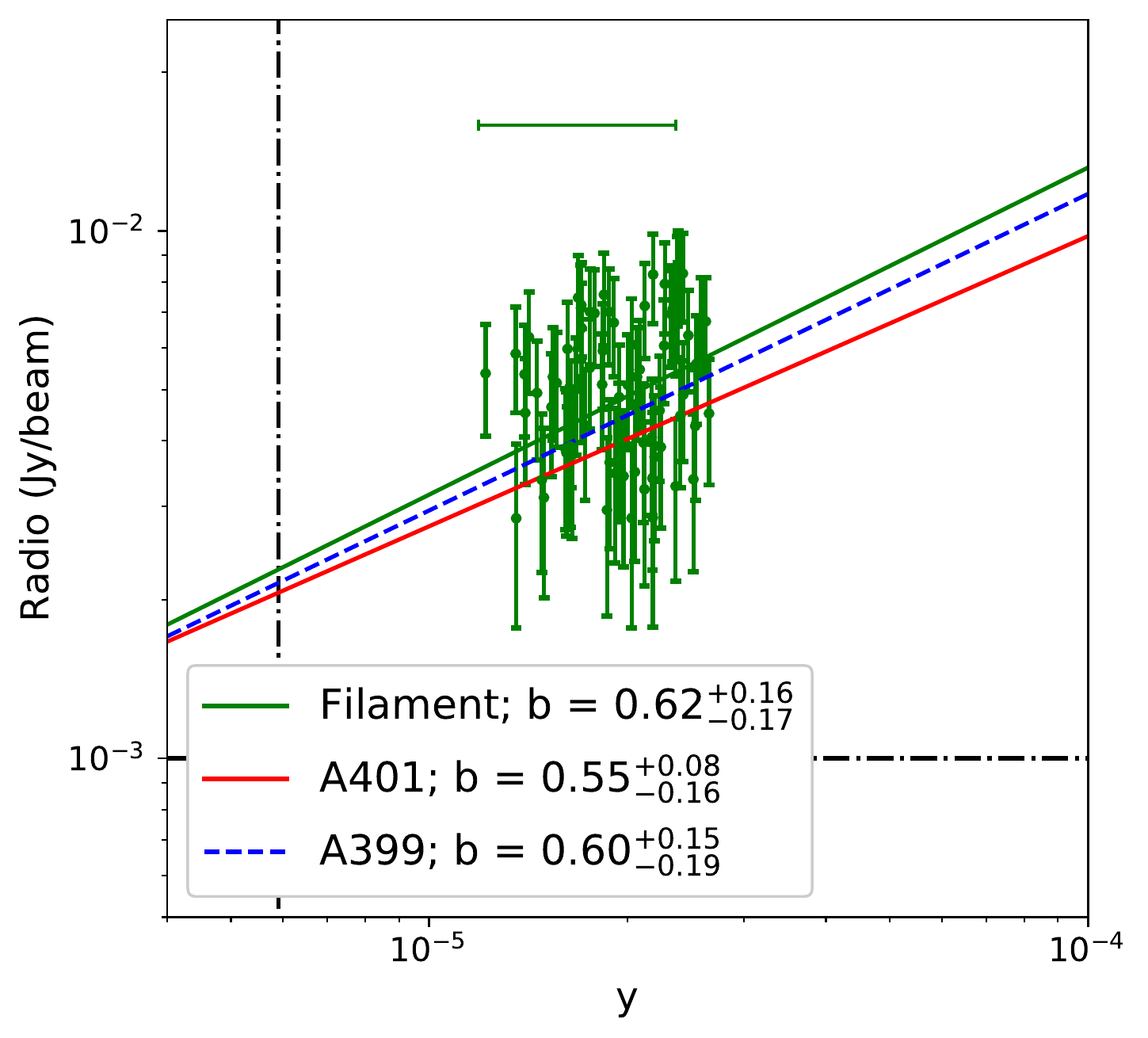}} 
    {\includegraphics[scale=0.4]{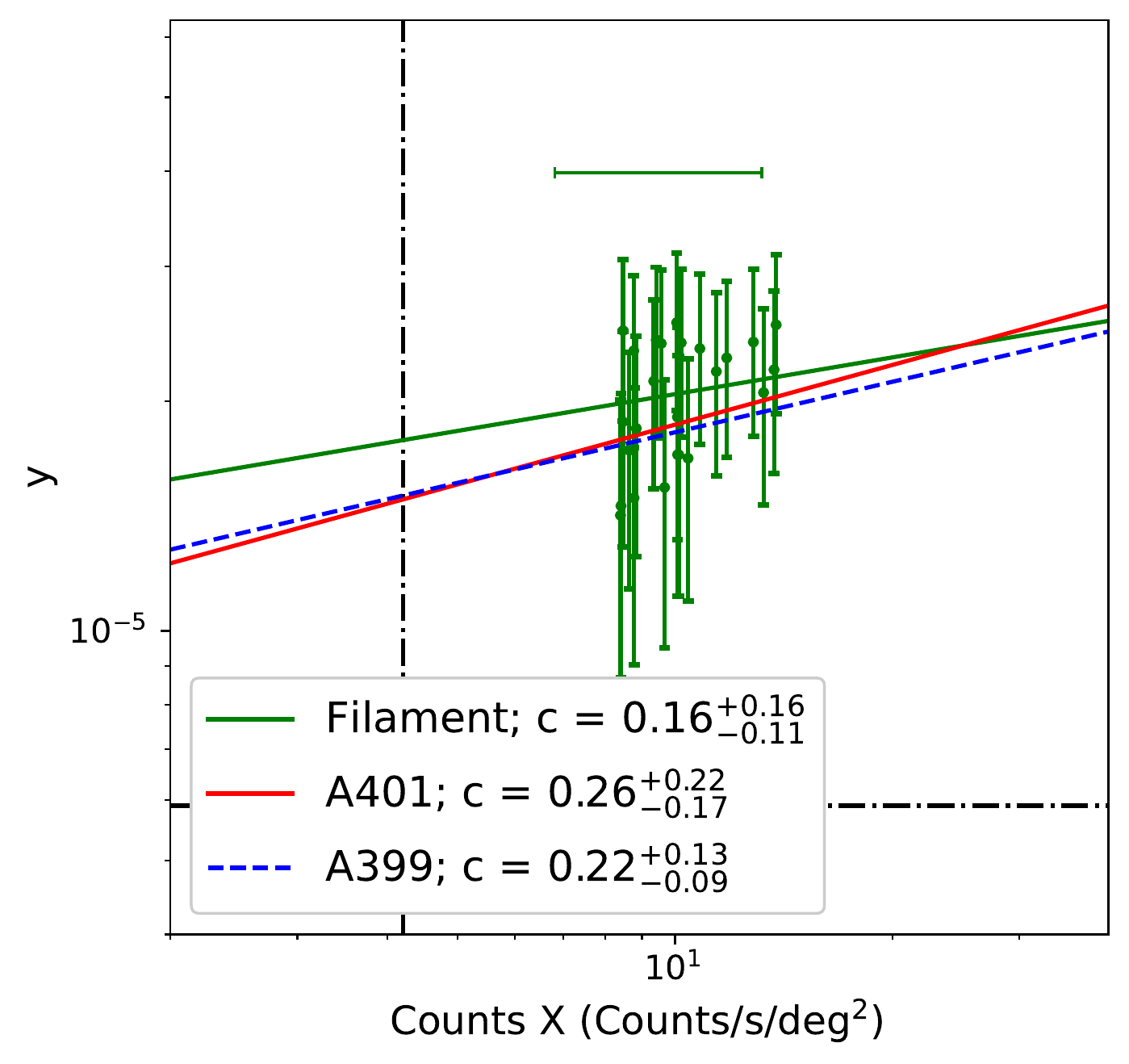}}
    \caption{\textbf{Left}: Correlation between the radio and X-ray count rate maps. \textbf{Middle}: Correlation between the radio map and the $y$ signal. \textbf{Right}: Correlation between the $y$ signal and the X-ray count rate map. The plotted points are only from the intercluster region. The figures show only the points that satisfy the condition $S/N > 2$ for each combination of the three maps. For each correlation, we report the best fit lines for A401 (red solid line) and A399 (blue dashed line), selecting only the points that satisfy the condition $2<S/N<5$. The horizontal and vertical dash-dotted lines indicate the noise level for each map we use. To convey the significance of the observed correlations, the average uncertainty for the distribution is indicated with a horizontal error bar in each panel.}
    \label{fig:Correlation_Bridge}
\end{figure*}

\subsection{Radio--X correlation}
\label{SubSec:X-radio}
We start by applying the grid to the LOFAR and \XMM maps. We extract the points shown in the upper panel of Fig.~\ref{fig:Correlation_SZ_radio}. For each region, we fit the data in the linear space with a power law that can be expressed in the log--log plane as
\begin{equation}
    \log \left( F_{\rm radio} \right)=a_{\rm radio-X} + d \, \log \left( F_{\rm X} \right),
    \label{eq:radio-x_corr}
\end{equation}
where $F_{\rm X}$ is the X-ray count rate per solid angle unit and $F_{\rm radio}$ the radio brightness in Jy/beam unit. We fit the model to the data, considering errors for both variables following a maximum likelihood approach as in \cite{logL}. The decision to fit the data in the linear space ensures that the uncertainties follow a normal distribution, a condition that may not always be satisfied in double log space. The same procedure is used in Secs.~\ref{SubSec:SZ-radio}, \ref{SubSec:SZ-X} and \ref{Appendix:Bridge}. The errors of the radio points are obtained by combining in quadrature the LOFAR thermal noise of the map, 1 mJy/beam, and the $15\%$ systematic error due to calibration, as reported in \cite{Govoni19}. To threshold the radio data, we use only the thermal noise of the map without considering the calibration error.
The radio--X scaling indices we find for A399 and A401 using the $S/N>2$ threshold are $d = 0.78 ^{+0.04} _{-0.04}$ and $d = 0.73 ^{+0.03} _{-0.03}$, respectively: a sublinear correlation for both the clusters.  
We estimate the level of correlation of the data in the two clusters by measuring the Pearson coefficients \citep[$\rho$ hereafter]{pearsonC} after linearizing the data. We find $\rho = 0.60$ and $\rho=0.79$ for A399 and A401, respectively. The relation in A399 is more scattered and less linearly correlated than in A401. Fit results are summarized in Table~\ref{tab:fit_params}. In Appendix~\ref{appendix:log_fit} we report the best-fit parameters when we fit the data in the double log scale.

In A399 \cite{sake04} reported a sharp X-ray edge in the eastern sector of the cluster: see fig. 4 of their paper. On the other hand, \cite{murgia2010} noted that the radio emission in that region does not drop off, indicating that the non-thermal components in this region of A399 extend beyond the thermal electrons population. The higher dispersion for A399 we note in Fig. \ref{fig:Correlation_SZ_radio} is possibly caused by this sharp X-ray edge. If we do not consider the boxes on the east sector of A399, the radio--X correlation is still sublinear but becomes steeper ($d=0.91 ^{+0.05} _{-0.05}$), more correlated ($\rho = 0.72$, so still smaller than A401), and consequently less scattered, indicating that our hypothesis is consistent with the data.

We also note that a cluster of points lies above the best fit trend for A401. A strong radio emission characterizes these points; we identify them as the compact radio emission in the north region of A401 visible in Fig. \ref{fig:Masks}. Even though this region may seem like a compact radio source, the analysis of its emission at 140 MHz at high spatial resolution and in the optical band indicates that no compact source is present at this spatial location. We therefore conclude that this emission most likely is diffuse and is associated with the intracluster medium. However, masking this region we find a scaling index of $0.71 ^{+0.03} _{-0.03}$, which is fully in agreement with the unmasked value, and a Pearson correlation coefficient of $\rho = 0.84$. See Appendix~\ref{appendix:masking_clusters} for further details about the fits masking the two regions of the clusters.

\subsection{Radio--SZ correlation}
\label{SubSec:SZ-radio}

In this section we apply the grid shown in Fig. \ref{fig:Masks} to the ACT+\textit{Planck} and LOFAR maps. The resulting points are shown in the middle panel of Fig. \ref{fig:Correlation_SZ_radio}. We fit the points with a power law, that in the log space results in 
\begin{equation}
    \log \left( F_{\rm radio} \right)= a_{\rm radio-SZ} + b \, \log \left( y \right),
    \label{eq:radio-SZ_corr}
\end{equation} 
where $F_{\rm radio}$ is the measured radio brightness and $y$ is the dimensionless SZ signal (equation~\ref{eq:tsz}). For $S/N>2$, we find that the scaling indices for A401 and A399 are $b = 1.47 ^{+0.08} _{-0.08} $ and $b = 2.83 ^{+0.08} _{-0.13}$. The Pearson correlation coefficients for the two clusters are respectively 0.74 and 0.60, similar to what was found in Sec. \ref{SubSec:X-radio}. Fit results for A399 and A401 are summarized in the first column of Table \ref{tab:fit_params}. The radio--SZ correlation in Fig. \ref{fig:Correlation_SZ_radio} seems to suggest that the slope of the correlation at the lowest flux densities is flatter than at high flux density. If we examine the scaling relations for A401 as we increase the $S/N$ threshold, however, the fitted slopes remain compatible at the $1 \sigma$ level. Thus there is no convincing evidence for a bias at the low flux densities in A401. The A399 data are much more scattered than for A401. As we will see in Sec. \ref{Appendix:Bridge}, the presence of scattering and a noise threshold could cause an apparent flattening of the data at low flux. The fact that we do not observe any significant  bias in A401 led us to consider all the available data in A399. However, we cannot exclude the possibility that A399 data suffer from an observational bias.

As we note in the caption of Fig. \ref{fig:Masks}, the SZ signal, like the X-ray brightness, has a sharp decline compared to the radio emission in the east sector of A399. If we do not consider the east sector of A399, we find $b = 3.02 ^{+0.06} _{-0.11}$ and $\rho = 0.68$. If we mask the northern region of A401, for the reasons given in Sec. \ref{SubSec:X-radio}, we find a slightly smaller scaling index, $1.42 ^{+0.08} _{-0.08}$, and as expected a Pearson coefficient closer to unity, $\rho = 0.80$.

\subsection{SZ--X correlation}
\label{SubSec:SZ-X}

We correlate the SZ map with X-ray data fitting
\begin{equation}
    \log \left(y\right) = a_{\rm SZ-X} + c \, \log \left( F_{\rm X} \right).
    \label{eq:SZ-X_corr}
\end{equation}
If we select only the data with $S/N>2$, we find $c= 0.50 ^{+0.02} _{-0.02}$ and $c= 0.51 ^{+0.03} _{-0.02}$ for A401 and A399 respectively. The data points and the best fit lines are shown in the bottom panel of Fig. \ref{fig:Correlation_SZ_radio}.
We note the relative lack of scatter as compared to the two correlations already discussed. Higher $S/N$ thresholds yield similar values.
The strong correlation between X-ray and SZ data is confirmed by the Pearson test. For A399 we find $\rho = 0.94$ while for A401 the parameter is 0.97; note that $\rho = 1.0$ means full correlation. Fit results for the SZ--X correlation are summarized in the third column of Table \ref{tab:fit_params}.

\begin{table*}
    \caption{Overview of the best-fit parameters using several $S/N$ cut-offs for the radio--SZ, radio--X, and SZ--X correlations in A401 and A399. For each pair of data sets, we report the Pearson correlation coefficient ($\rho$). Note that best-fit parameters are expressed in the double log scale format.}
    \centering
    \begin{center}
    \begin{tabular}{cccc|ccc|ccc}
        \multicolumn{10}{c}{ \textbf{$S/N>2$} } \\ 
         & \multicolumn{3}{c}{ \textbf{radio--SZ} $\left( \mathrm{equation}~\ref{eq:radio-SZ_corr} \right)$ } & \multicolumn{3}{c}{ \textbf{radio--X} $\left( \mathrm{equation}~\ref{eq:radio-x_corr} \right)$} & \multicolumn{3}{c}{ \textbf{SZ--X} $\left( \mathrm{equation}~\ref{eq:SZ-X_corr} \right)$ } \\
        \hline
        \hline
        \multicolumn{1}{c|}{Region} & a$_{\rm radio-SZ}$ & b & $\rho$ & a$_{\rm radio-X}$ & d & $\rho$ & a$_{\rm SZ-X}$ & c & $\rho$\\
        \hline
        \multicolumn{1}{c|}{A401} & $ 4.51 ^{+0.52} _{-0.23} $ & $1.47 ^{+0.08} _{-0.08}$ & 0.74 & 
        $ -3.22 ^{+0.06} _{-0.06}$ & $0.73 ^{+0.03} _{-0.03}$  & 0.79 & 
        $ -5.26 ^{+0.04} _{-0.04}$ & $0.50 ^{+0.02} _{-0.02}$ & 0.97\\
        
        \multicolumn{1}{c|}{A399} & $ 11.14 ^{+0.59} _{-0.32} $ & $2.83 ^{+0.08} _{-0.13}$  & 0.60 & 
        $ -3.12 ^{+0.07} _{-0.06}$ & $0.78 ^{+0.04} _{-0.04}$  & 0.60 & 
        $ -5.28 ^{+0.04} _{-0.04}$ & $0.51 ^{+0.03} _{-0.02}$ & 0.94\\

        \hline
    \end{tabular}
    \end{center}
    
    \begin{center}
    \begin{tabular}{cccc|ccc|ccc}
        \\
        \multicolumn{10}{c}{ \textbf{$S/N>3$} } \\ 
        \hline
        \hline
        \multicolumn{1}{c|}{Region} & a$_{\rm radio-SZ}$ & b & $\rho$ & a$_{\rm radio-X}$ & d & $\rho$ & a$_{\rm SZ-X}$ & c & $\rho$\\
        \hline
        \multicolumn{1}{c|}{A401} & $ 4.71 ^{+0.65} _{-0.25} $ & $1.52 ^{+0.09} _{-0.09}$ & 0.71 & 
        $ -3.22 ^{+0.07} _{-0.06}$ & $0.73 ^{+0.03} _{-0.03}$  & 0.76 & 
        $ -5.24 ^{+0.04} _{-0.04}$ & $0.49 ^{+0.02} _{-0.02}$ & 0.97\\
        
        \multicolumn{1}{c|}{A399} & $ 12.41 ^{+0.71} _{-0.34} $ & $3.11 ^{+0.09} _{-0.15}$ & 0.51 & 
        $ -3.29 ^{+0.10} _{-0.09}$ & $0.88 ^{+0.06} _{-0.06}$  & 0.57 & 
        $ -5.25 ^{+0.05} _{-0.05}$ & $0.49 ^{+0.03} _{-0.03}$ & 0.93\\
        
        \hline
    \end{tabular}
    \end{center}

    \begin{center}
    \begin{tabular}{cccc|ccc|ccc}
        \\
        \multicolumn{10}{c}{ \textbf{$S/N>5$} } \\
        \hline
        \hline
        \multicolumn{1}{c|}{Region} & a$_{\rm radio-SZ}$ & b & $\rho$ & a$_{\rm radio-X}$ & d & $\rho$ & a$_{\rm SZ-X}$ & c & $\rho$\\
        \hline
        \multicolumn{1}{c|}{A401} & $ 4.88 ^{+0.94} _{-0.29}$ & $1.56 ^{+0.12} _{-0.11}$ & 0.55 & 
        $ -3.16 ^{+0.09} _{-0.08}$ & $0.71 ^{+0.04} _{-0.04}$  & 0.66 & 
        $ -5.22 ^{+0.05} _{-0.05}$ & $0.49 ^{+0.02} _{-0.02}$  & 0.98\\
        
        \multicolumn{1}{c|}{A399} & $ 14.75 ^{+0.24} _{-0.26} $ & $3.69 ^{+0.05} _{-0.11}$  & 0.44 & 
        $ -3.38 ^{+0.15} _{-0.13}$ & $0.93 ^{+0.09} _{-0.08}$  & 0.46 & 
        $ -5.12 ^{+0.08} _{-0.08}$ & $0.43 ^{+0.05} _{-0.04}$  & 0.94\\
        
        \hline
    \end{tabular}
    \end{center}
    
    \label{tab:fit_params}
\end{table*}

\subsection{Correlation in the filament}
\label{Appendix:Bridge}

\begin{table*}
    \caption{Overview of the best-fit parameters in the filament using the threshold $S/N>2$ and in the two clusters when only the points which satisfy the condition $2<S/N<5$ are selected. For each pair of data sets, we report the Pearson correlation coefficient ($\rho$). Note that best-fit parameters are expressed in the double log scale format.}
    \centering
    \begin{center}
    \begin{tabular}{cccc|ccc|ccc}
        \multicolumn{10}{c}{ \textbf{$S/N>2$} } \\ 
         & \multicolumn{3}{c}{ \textbf{radio--SZ} $\left( \mathrm{equation}~\ref{eq:radio-SZ_corr} \right)$ } & \multicolumn{3}{c}{ \textbf{radio--X} $\left( \mathrm{equation}~\ref{eq:radio-x_corr} \right)$} & \multicolumn{3}{c}{ \textbf{SZ--X} $\left( \mathrm{equation}~\ref{eq:SZ-X_corr} \right)$ } \\
        \hline
        \hline
        \multicolumn{1}{c|}{Region} & a$_{\rm radio-SZ}$ & b & $\rho$ & a$_{\rm radio-X}$ & d & $\rho$ & a$_{\rm SZ-X}$ & c & $\rho$\\
        \hline
        \multicolumn{1}{c|}{Filament} & $0.60 ^{+1.91} _{-0.37} $ & $0.62 ^{+0.16} _{-0.17} $ & 0.15&
        $ -2.60 ^{+0.35} _{-0.22} $ & $0.34 ^{+0.31} _{-0.26}$ & -0.08 & 
        $ -4.85 ^{+0.12} _{-0.13} $ & $0.16 ^{+0.16} _{-0.11}$ & 0.45 \\
        \hline
    \end{tabular}
    \end{center}
    
    \begin{center}
    \begin{tabular}{cccc|ccc|ccc}
        \\
        \multicolumn{10}{c}{ \textbf{$2<S/N<5$} } \\ 
        \hline
        \hline
        \multicolumn{1}{c|}{Region} & a$_{\rm radio-SZ}$ & b & $\rho$ & a$_{\rm radio-X}$ & d & $\rho$ & a$_{\rm SZ-X}$ & c & $\rho$\\
        \hline
        \multicolumn{1}{c|}{A401} & $ 0.19 ^{+0.64} _{-0.36} $ & $0.55 ^{+0.08} _{-0.16}$ & 0.35 & 
        $ -2.58 ^{+0.17} _{-0.19}$ & $0.16 ^{+0.20} _{-0.12}$ & 0.31 & 
        $ -4.99 ^{+0.26} _{-0.20}$ & $0.26 ^{+0.22} _{-0.17}$ & 0.54\\
        
        \multicolumn{1}{c|}{A399} & $ 0.47 ^{+1.78} _{-0.38} $ & $0.60 ^{+0.15} _{-0.19}$  & 0.55 & 
        $ -2.58 ^{+0.14} _{-0.16}$ & $0.19 ^{+0.17} _{-0.11}$ & 0.41 & 
        $ -4.96 ^{+0.11} _{-0.12}$ & $0.22 ^{+0.13} _{-0.09}$  & 0.56\\
        
        \hline
    \end{tabular}
    \end{center}
    
    \label{tab:fit_params_fil}
\end{table*}

This subsection investigates the filament between the two clusters. In Fig. \ref{fig:Correlation_Bridge} we show the three correlations extracted from the intercluster region using an $S/N>2$ threshold. We find $b = 0.62 ^{+0.16} _{-0.17}$, $d = 0.34 ^{+0.31} _{-0.26}$, and $c = 0.16 ^{+0.16} _{-0.11}$ respectively for the radio--SZ, radio--X, and SZ--X correlations. These relations are flatter than those of the two clusters and are summarized in Table \ref{tab:fit_params_fil}. Moreover, we note that the three scaling indices are characterized by a large uncertainty, making them compatible with zero at the 2--3$\sigma$ level. We also note that especially for the radio--X and SZ--X correlations, the X-ray points could be effectively a single point due to the large uncertainties in the X-ray intensity. The Pearson correlation coefficients show a negligible correlation for the radio--SZ and radio--X comparisons. They are $\rho = 0.15$ and $\rho = -0.08$, while for the spatial correlation between SZ and X-ray, we find $\rho = 0.45$. These values are considerably smaller than those found in the individual clusters of this system, and the correlations in Fig. \ref{fig:Correlation_Bridge} appear to be less significant than in A399 and A401. In order to estimate the probability that two uncorrelated populations produce a data set described by the same Pearson coefficients observed in the filament, the $p$-value associated with each correlation is computed, resulting in 30\% for the radio–SZ, 71\% for the radio-X, and 3\% for the SZ-X. The probability that the radio--SZ and radio--X correlations arise from an uncorrelated data set is not negligible, although a filament of hot gas between the two clusters has been detected in all the maps we use.

One might think that we observe only the brightest part of the filament due to the low density of the intercluster medium and/or the projection of the system on the plane of the sky, as already has been observed by \cite{hincks22}, and that this in turn biases the resulting scaling index and amplitude of the fits. To understand if the flatter relations we find for the radio--X, radio--SZ, and SZ--X comparisons are due to some observational bias, we repeat the study on the two clusters, focusing our attention on the faintest regions. 
We examine the correlations in the clusters selecting only data between $S/N > 2$ and $S/N < 5$. This means that we are using the boxes at the edge of the clusters, which have a brightness comparable with what we see in the filament. These regions are indicated in Fig. \ref{fig:Masks} with the thinnest outlines. Fit results for these faint regions of the clusters are summarized at the bottom of Table~\ref{tab:fit_params_fil}, and the corresponding best fit lines, together with the scaling relation slopes, are also shown in Fig.~\ref{fig:Correlation_Bridge}. We note the consistency between the best fit lines for the three observables.
The Pearson correlation coefficients reside in the range 0.3--0.5 for all the cases, resulting less linearly correlated than the data reported in Table~\ref{tab:fit_params}.

The best fit parameters we extract using only the faintest regions in the two clusters are $1\sigma$ compatible with the ones in the filament. This suggests that the flatter relations in the filament are very likely due to an observational bias. Among the possible biases, we cite the fact that we are detecting only the brightest regions of the filaments and that there is likely intrinsic scatter in the data. Intrinsic scatter would affect the low $S/N$ data more, resulting in a possible deviation of the scaling index of the relation from the true value when only low significance data are examined. To demonstrate this possibility, we perform some simple simulations that include intrinsic scatter in the presence of a pure noise threshold of $2\sigma$ and reproduce the flatter correlation found in faint regions of the filament and clusters. In more detail, first we generate points following the cluster's best-fit scaling relations (see Table \ref{tab:fit_params}), then we scatter the points by combining the data's noise and an additional term for intrinsic scatter, and finally we threshold the data and fit the points satisfying the $S/N>2$ condition. After we apply the noise threshold, the points near the $2\sigma$ noise threshold systematically lie above the best-fit scaling relation we used to generate the points themselves. This results in a flatter correlation, and thus we conclude that there is no evidence for different correlation properties between faint and bright regions for the three pairs of observables we study. Deeper, lower-noise observations of this system or systems like it will be required to obtain statistically reliable correlations in the faint regions and thereby determine if different physical mechanisms operate in the cluster and filament regions. However, this study is beyond the scope of the paper, so we will not further investigate this aspect.

\section{Further considerations on the correlations}
\label{Sect:Interpretations}

The results we have presented in Sec. \ref{Sect:Correlation} can be used to infer properties of the two clusters, such as the relations between thermal and non-thermal components, the temperature radial profile, and the slope of the radial dependence of the magnetic field strength, $B(r)$. This section will focus on the former. We will compare our results with previous work on galaxy clusters hosting radio halos. The discussion of the magnetic field is left to Sec. \ref{Sec:Magnetic_Field_Slope}.

\subsection{Radio--X correlation}
\label{SubSec: radio-X interp}

The characterization of the radio--X correlation in galaxy clusters hosting radio halos has been done in about ten systems over the last two decades. \cite{Govoni01} analyzed for the first time the local radio--X correlation for four galaxy clusters. They found a linear relation for two of them (Abell 2255, confirmed by \citealt{Botteon20}, and Abell 2744, in which \citealt{Rap21}, by contrast, found a sublinear relation) and a sublinear trend for the other two (the Coma cluster and Abell 2319). \cite{Bonafede21} studied the radio--X correlation in the radio bridge of the Coma cluster, finding a correlation parameter of $0.5\pm0.1$. Recently \cite{vacca22} found a sublinear scaling index in Abell 523 at 144 MHz despite the weak linear correlation in the log--log plane, while \cite{cova19} detect a broad intrinsic scatter between these two quantities at 1.4 GHz. On the other hand, \cite{hoang19}, studying Abell 520, discovered only a tentative trend between $Chandra$ data and three radio maps at different frequencies. Similarly, \cite{Shim14} found no significant correlation in the radio halo of the Bullet cluster (1E 0657-55.8) at the sensitivity limit of their observations. \cite{riseley2022}, comparing radio and X-ray brightness for the radio halo in Abell 3266, reported a linear scaling relation ($0.99 \pm 0.07$) while 
\cite{Rajp18} found a log--log slope of $1.25 \pm 0.16$ in the 1RXS J0603.3+4214 cluster.

Previous work has been done at 140\,MHz and 1.4\,GHz, but as shown in \cite{Rajpurohit2021}, and confirmed in \cite{Rap21}, this should not affect the results within the uncertainties. Most of the radio halos studied in the past show a linear or sublinear radio--X relation; however, the number of well-studied systems is still small, and there is not yet a clear view of the matter.

In this work, we studied the radio halos A399 and A401, for which high-angular resolution multi-wavelength data ($\sim$ arcminute) are available. We find that a sublinear radio-X correlation characterizes the two systems. As shown in \cite{Govoni01}, a linear relation between the radio surface brightness and X-ray count rate per solid angle unit translates to
\begin{equation}
N_{0} B^{(\delta+1)/2} \propto n_{\mathrm{e}} ^2 (k T)^{1/2},
\label{eq:X-radio_relation}
\end{equation}
assuming that the number density of relativistic electrons with energy between $\epsilon$ and $\epsilon + d \epsilon$ is $n_{e} d \epsilon = N_{0} \epsilon^{-\delta} d \epsilon$, and where $B$ is the magnetic field and $T$ is the gas temperature. Note that equation~(\ref{eq:X-radio_relation}) assumes pure bremsstrahlung emission for the X-ray counterpart, which is a good approximation only when the temperature is greater than a few keV. Assuming that the cluster is isothermal, a sublinear radio--X correlation suggests that the radial decline of magnetic fields and non-thermal particles is slower than the square of the thermal gas density, as we have found for both the clusters, or the opposite in the superlinear regime.

As pointed out in \cite{Govoni01} the relation between radio and X-ray brightness could be used to infer which model is most likely responsible for the relativistic electrons generally found in radio halos, including those in A399 and A401. These authors showed that re-acceleration scenarios predict a linear or sublinear radio–X correlation, while with injection models a linear or superlinear correlation depending on the magnetic field strength is anticipated. 
We conclude that the data analysed in this section indicate a preference for the re-acceleration scenario; however, further analysis and a larger statistical sample of the radio--X correlations in other radio halos are of fundamental importance before drawing a firm conclusion on the origin of non-thermal electrons.

\subsection{SZ--X correlation}
\label{SubSec: SZ-X interp}

The correlation between SZ and X-ray indicates that the ICM in A399 and A401 can be well described with an isothermal profile. Indeed for these two clusters, we find a scaling index compatible with $\simeq0.5$. The X-ray thermal bremsstrahlung is proportional to $F_{\rm X} \propto n_{e}^{2} T^{\frac{1}{2}}$, while the SZ effect scales as $y_\textsc{sz} \propto n_{e} T$. If the gas temperature is constant over the cluster volume or almost constant with respect to density variations, the SZ--X correlation is proportional to $n_{e}^{0.5}$, which is consistent with the scaling indices we observe in A399 and A401.

However, we also note that the scaling index of the SZ--X correlation in A399 is smaller when the $S/N$ threshold is $>5$, although it remains compatible at the 2$\sigma$ level with the value we expect for an isothermal medium. This small deviation from the value of 0.5 might be an indication of weak temperature gradients inside A399, as already suggested by \cite{sake04} and \cite{Bourdin08}.

\subsection{Radio--SZ correlation}
\label{SubSec: radio-sz interp}

It is well known that there is a global relation between the integrated radio  emission and the integrated Compton--$y$ parameter in clusters of galaxies hosting radio halos. \citet{basu12}, B12 hereafter, discovered a correlation between cluster radio power at 1.4 GHz and the Compton-$y$ parameter integrated up to $5 R_{500}$ with a log--log slope of $1.88 \pm 0.24$. 
B12 also considered another sample of clusters observed at 327\,MHz; the trend of the correlation is consistent with result at 1.4\,GHz. \cite{weeren21}, W21 hereafter, determined the radio--SZ scaling relation for a sample of radio halos using LOFAR 150 MHz and Compton $Y_{R500}$ data from \cite{Planck2016}. W21 found a steeper log--log slope with respect to B12, equal to $3.32 \pm 0.65$ using the orthogonal regression algorithm \citep{ortho96}. However, \cite{weeren21} also found that the log--log slope is significantly affected by the fitting method; adapting the same model to the same data but using three different techniques, W21 only found a $\gtrsim 1 \sigma$ compatibility between the resulting scaling relations. In fitting the integrated radio power $P$ vs. $Y_{R500}$, where $P$ and $Y_{R500}$ are the independent and the dependent variables, a scaling index of $2.1 \pm 0.67$ \citep{weeren21}, compatible with B12, was reported. \cite{Planck13_Coma}, PLA13 hereafter, studied the Coma cluster using the new \textit{Planck} SZ map with an angular resolution of 10$^{\prime}$ and radio data at 352 MHz, finding for the first time a point-by-point correlation between radio and SZ brightnesses inside a cluster. A quasi-linear relation with a scaling index of $0.92 \pm 0.04$, fitting SZ versus radio brightness was reported, which corresponds to a $\sim$1.1 reversing of the relation between radio brightness and SZ, as we have also done here.

In this work, we correlate radio and SZ data in the inner part of the galaxy clusters like PLA13, but with a significantly higher resolution of $1.65^{\prime}$. The radio--SZ relations we find for A399 and A401 are much steeper than the PLA13 results. The main differences are the angular resolution and the frequency of the radio map. While we believe that the different frequencies cannot be the only cause of the discrepancy, the different resolutions may solve the tension. Our higher angular resolution data allows us to define a fine grid that excludes compact sources. Residual sources could still affect the brightness in the boxes at the edge of the clusters. An increase of the brightness at the cluster's edge due to source contamination would result in a flatter relation in the log--log plane. This effect has been noticed in \cite{Bonafede_2022}. The latter authors have reanalyzed the radio--SZ relation in the Coma cluster, with a better resolution than PLA13, finding a steeper slope. The scaling indices we present in this work for A339 and A401 are $1 \sigma$ compatible with the global trend discovered by \cite{weeren21}, which used the same radio frequency. Moreover, for A401, our result is $2 \sigma$ compatible with B12 results, while for A399 we report a tension slightly smaller than $4 \sigma$ with respect to B12.
However, our work indicates that the relation between the non-thermal components and thermal pressure is also verified on sub-cluster scales.  

The radio--SZ relation we find for A399 and A401 can be written as
\begin{equation}
    N_{0} B^{(\delta+1)/2} \propto \left ( n_{e} k T \right )^{b} , 
    \label{eq:radio-SZ}
\end{equation}
where $b$ is the power law index reported in Table \ref{tab:fit_params}. We note that the radio--SZ scaling index for A401 is around twice the one we extract from radio and X-ray data. This result is in line with what we expect for an almost isothermal medium, or slowly variable gas temperature along the radial direction, considering that the radio--X correlation can be written as
\begin{equation}
    N_{0} B^{(\delta+1)/2} \propto \left ( n_{e} ^2 (k T)^{1/2} \right )^{d}.
    \label{eq:radio-X}
\end{equation}
If we compare radio--SZ (equation~\ref{eq:radio-SZ}) and radio--X (equation~\ref{eq:radio-X}) correlations and assume that the gas can be approximated as isothermal, we expect $b = 2d$. Hence, from the radio--X correlation in A401, we would expect a radio--SZ scaling index of ${\sim}1.5$, which fully agrees with the radio--SZ scaling index we measure in A401. The same trend is also verified when we mask the up-scattered radio points (see Appendix~\ref{appendix:masking_clusters}). In the case of A399, if the assumption of an isothermal ICM were correct from the radio--X comparison, we would expect a radio--SZ scaling index of $\sim 1.60$. This value is clearly incompatible with the one we measure comparing radio and SZ data, $2.83 ^{+0.08} _{-0.13}$, and masking the east sector of A399 does not alleviate the tension. In Sec. \ref{SubSec: SZ-X interp} we find that an isothermal profile can provide a good global description of A399. However, we do not exclude the presence of a slight temperature gradient in the ICM responsible for the drift of the SZ--X scaling index when increasing the $S/N$ threshold. \cite{Bourdin08} classified A399 as an `irregular' cluster due to past merger activities within the cluster; our findings seem to confirm this conclusion using multi-frequency data instead of only X-ray. The Pearson correlation test for the radio--X and radio--SZ correlations indicates a much more disturbed cluster morphology with respect to A401. Note that when the east sector of A399 is masked, the levels of correlation in A399, measured with the Pearson test, remain smaller than the ones in A401. Deeper high resolution data would help investigate individual cluster sectors further and eventually, better understand A399 morphology.

\section{Radial dependence of the magnetic field}
\label{Sec:Magnetic_Field_Slope}

\begin{figure*}
    \centering
    {\includegraphics[scale=0.62]{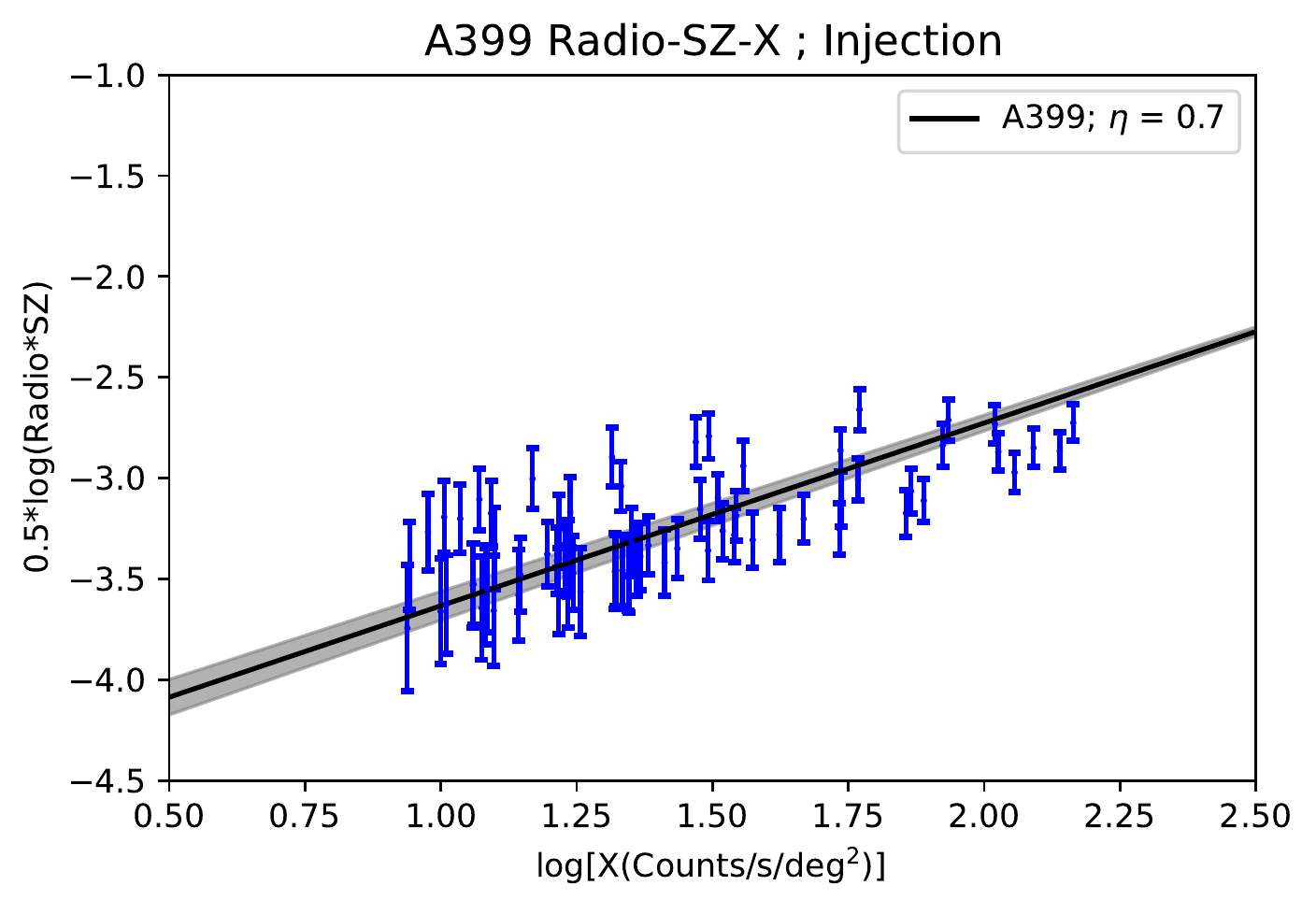}}
    {\includegraphics[scale=0.62]{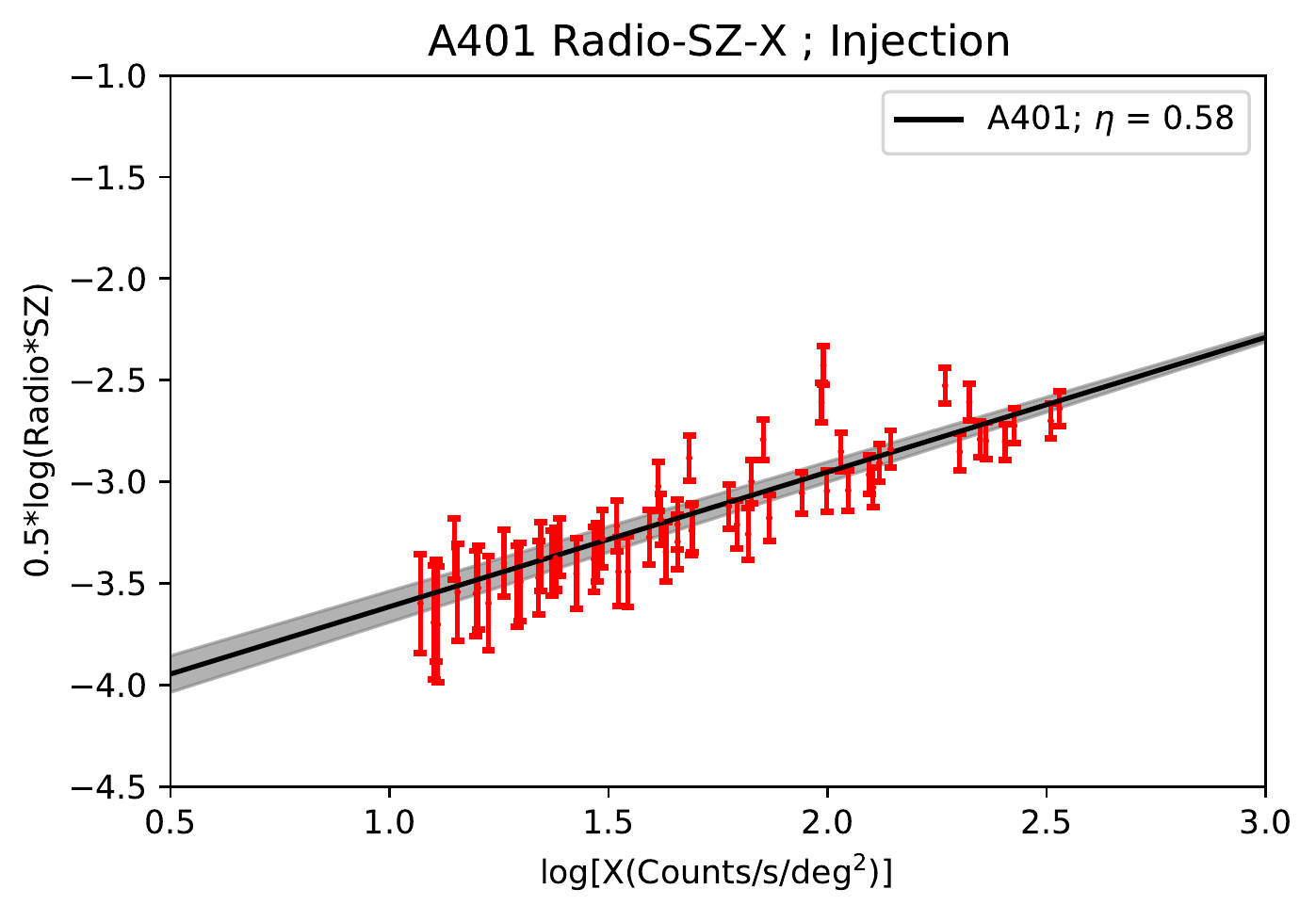}}
    \caption{\textbf{Left}: A399 radio--X correlation points are shown in blue. The black line shows the fitted points using SZ, radio and X-ray data. \textbf{Right}: The same for Abell 401. The measured points are shown in red. Grey shadows indicate $\eta$--$1 \sigma$ confidence band.}
    \label{fig:Triple_Correlation}
\end{figure*}

The data set presented in Sec. \ref{Sect:data} and analyzed in Secs.~\ref{Sect:Correlation} and \ref{Sect:Interpretations} can be used to study the magnetic field properties in the two clusters. In this section, starting from theoretical assumptions and using radio, SZ, and X-ray data, we derive information about magnetic fields in A399 and A401.

There are theoretical \citep[e.g.,][]{Dolag2002, donnert18} and observational \citep[e.g.,][]{dolag01, murgia2004} arguments indicating that magnetic fields inside galaxy clusters must decrease with the increase of the radial distance from the cluster center scaling as a function of the electron density. In this work, we assume that the magnetic field in galaxy clusters hosting radio halos follows the following radial profile
\begin{equation}
    \langle B(r) \rangle = B_{0} \left( \frac{n_{\mathrm{e}}(r)}{n_{0}} \right)^{\eta},
    \label{eq: Magnetic_field}
\end{equation}
where $B_{0}$ and $n_{0}$ are the central magnetic field and the central thermal electron density, respectively, $n_{\mathrm{e}}(r)$ is the electron density at a given radius $r$ from the cluster center, and $\eta$ is the scaling index of the magnetic field, expected to be in the range 0.3--1 \citep[e.g.,][and references therein]{Murgia09}. A positive value of $\eta$ combined with a radially decreasing electron density results in a radial decrease of the magnetic field.
The parameter $\eta$ is related to the cluster's gravitational collapse mechanism and the overall amplification process of $B$ \citep[e.g.,][]{vazza18}. Under the hypothesis of conservation of mass and magnetic field flux, in the case of a frozen-in magnetic field during the compression of a spherical cloud we expect $\eta=2/3$ \citep{Ferretti04}. If the magnetic field energy scales as the thermal energy, $\eta$ is predicted to be 0.5 \citep[e.g.,][]{bonafede10}. Different values can be obtained in different non-spherical geometries.

Assuming an isothermal $\beta$--model \citep{Cav&Fusco76}, equation~(\ref{eq: Magnetic_field}) becomes
\begin{equation}
    \langle B(r) \rangle = B_{0} \left( 1+ \frac{r^{2}}{r_{\mathrm{c}}^{2}} \right)^{-\frac{3}{2} \beta \eta},
    \label{eq: Magnetic_field_betaM}
\end{equation}
where $\beta$ is the slope of the density profile and $r_{\mathrm{c}}$ is the core radius. For a $\beta$--model, the brightness profile for X-ray emission and SZ signal can be written respectively as \citep[e.g.,][]{Hughes_1998} 

\begin{equation}
    I_\textsc{x} (r) \propto T^{1/2} \left( 1 + \frac{r^2}{r_{\mathrm{c}} ^{2}} \right)^{-3 \beta + 0.5}
    \label{eq:X_bright}
\end{equation}
\begin{equation}
    I_\textsc{sz} (r) \propto T \left( 1 + \frac{r^2}{r_{\mathrm{c}} ^{2}} \right)^{-3/2 \beta + 0.5}.
    \label{eq:SZ_bright}
\end{equation}
In the following, we will derive the radio brightness profile and the index $\eta$ for the injection and re-acceleration scenarios.

\subsection{Radial Dependence of the Magnetic Field in the Injection Model}
\label{SubSec:Injection_model}

In the first scenario, we assume that cosmic ray electrons are continuously injected into the ICM. If this electron population is described with a power-law distribution between a minimum and maximum energy range, then the brightness profile for radio emission under the equipartition conditions is,
\begin{equation}
    I_{\rm radio} (r) = I_{\rm radio, 0} \left( 1 + \frac{r^2}{r_{\mathrm{c}} ^{2}} \right)^{-6 \beta \eta + 0.5} ,
    \label{eq:radio_bright_injection}
\end{equation}
considering that, under these conditions, the local radio brightness is proportional to $B^4$ \citep{Murgia09}.

Given the high-resolution data we have, we can combine the following two relations to estimate $\eta$ in this injection scenario
\begin{equation}
    \frac{I_{\rm radio}(r)}{I_\textsc{sz}(r)} \propto T^{-1} \left( 1 + \frac{r^2}{r_{\mathrm{c}} ^{2}} \right)^{-3 \beta \left( 2 \eta - 0.5 \right)}
    \label{eq:radio-sz}
\end{equation}

\begin{equation}
    \frac{I_{\rm radio}(r)}{I_\textsc{x}(r)} \propto T^{-0.5} \left( 1 + \frac{r^2}{r_{\mathrm{c}} ^{2}} \right)^{-6 \beta \eta + 3 \beta} .
    \label{eq:radio-x}
\end{equation}

For the first time here we can use the radio--SZ spatial correlation in galaxy clusters (equation~\ref{eq:radio-sz}) to remove the temperature dependence in equation~(\ref{eq:radio-x}) and then extract $\eta$ both for A399 and A401. Combining the previous two relations, we find
\begin{equation}
    \frac{I_{\rm radio}(r)}{I_\textsc{x}(r)} \propto \left( \frac{I_{\rm radio}(r)}{I_\textsc{sz}(r)} \right)^{\frac{1}{2}} \left( 1 + \frac{r^2}{r_{\mathrm{c}} ^{2}} \right)^{-3 \beta \left( \eta - \frac{3}{4} \right)}.
    \label{eq:triple_corr_lin}
\end{equation}
This expression can be used to estimate $\eta$ and unlike equations~\ref{eq:radio-sz} and \ref{eq:radio-x} is independent of gas temperature. We adopt the core radii and slopes measured by \cite{hincks22} using a spherical $\beta$--model to fit the two clusters. These authors found $\beta _{A401} = 0.83 \pm 0.06$ and $r_{c, A401} = 2.4 \pm 0.3 ^{\prime}$ for A401, while for A399 they found $\beta _{A399} = 0.80 \pm 0.09$ and $r_{c, A399} = 2.9 \pm 0.6 ^{\prime}$. To estimate $\eta$ we fit equation~(\ref{eq:triple_corr_lin}) in the form
\begin{equation}
    \begin{split}
    \frac{1}{2}\log \left( I_{\rm radio}(r) \cdot I_\textsc{sz}(r) \right) = a +\log \left( I_\textsc{x}(r) \right)  + \\ -3 \beta \left( \eta - \frac{3}{4} \right) \log\left( 1 + \frac{r^2}{r_{\mathrm{c}} ^{2}} \right) ,
    \end{split}
\end{equation}
where $a$ is an offset term. The results for A399 and A401 are $\eta _{A399} =  0.70 \pm 0.04$ and $\eta _{A401} = 0.58 \pm 0.03 $, respectively. The fits to the correlations for the injection model, using all three data sets together, are shown in Fig. \ref{fig:Triple_Correlation}. Note that these measurements of the magnetic field scaling indices depend on the assumption of a circular $\beta$-profile for the two clusters. To properly quantify how the values we assumed for $r_{c}$ and $\beta$ affect our results, we tested the cases where we set these two parameters at the extremes of the 1$\sigma$ confidence levels reported in \cite{hincks22}. We find that this creates an additional uncertainty of 0.03 and 0.01 for $\eta_{401}$ and $\eta_{399}$, respectively. These uncertainties are reported in Table \ref{tab:eta} inside the brackets.

The advantage of fitting the three data sets at the same time is clear when examining the Pearson coefficient of the quantities shown in Fig. \ref{fig:Triple_Correlation}. For A401, we find $\rho = 0.90$, while for A399, we find $\rho = 0.74$ (0.83 when we mask the east sector). Both Pearson values reveal a more significant linear relation than the ones we extracted in Sec. \ref{Sect:Correlation} for the radio--X and radio--SZ correlations. As expected, by introducing a third data set, we reduce the intrinsic scatter of the two double correlations, making the estimation of the scaling index $\eta$ more robust. If we use only the radio and X-ray data (equation~\ref{eq:radio-x}) to extract the scaling index of the magnetic field we would find $\eta _{A399} ^{X,R} =  0.54 \pm 0.06 $ and $\eta _{A401} ^{X,R} = 0.44 \pm 0.05$. However, the lower Pearson coefficients we found in the radio--X correlations for A401 and A399 (0.73 and 0.6) indicate a poorer quality of the fit compared to using the three maps simultaneously. We conclude that the radial dependence of the magnetic field estimation with the triple data set is more reliable than the values we can derive using the ratio between radio and X-ray brightness or only the radio brightness radial profile.

\subsection{Radial Dependence of the Magnetic Field in the Re-acceleration model}
\label{SubSec:Re-acc}

The observed radio brightness profile of the galaxy clusters can also be described with a second model assuming that existing electrons are continuously re-accelerated and balance radiative losses. In a simplified version of this model, under the equipartition conditions, the radio brightness can be written as
\begin{equation}
    I_{\rm radio} (r) = I_{0} \left( 1 + \frac{r^{2}}{r_{\mathrm{c}} ^{2}} \right)^{-5.25 \beta \eta + 0.5}.
\end{equation}
Note that in this scenario, under perfect equipartition between the particles' energy density and magnetic field, the local radio brightness scales as $B^{3.5}$. See \cite{Murgia09} for further details about the re-acceleration scenario.

Combining the previous equation with equation~(\ref{eq:X_bright}) \& (\ref{eq:SZ_bright}), in the re-acceleration scenario, we obtain the following relation between radio, X-ray, and SZ brightness profiles
\begin{equation}
    \frac{I_{\rm radio}(r)}{I_\textsc{x}(r)} \propto \left( \frac{I_{\rm radio}(r)}{I_\textsc{sz}(r)} \right)^{\frac{1}{2}} \left( 1 + \frac{r^2}{r_{\mathrm{c}} ^{2}} \right)^{ -\frac{21}{8}\beta \left( \eta - \frac{18}{21} \right)},
\end{equation}
from which we can extract $\eta$ as we did in Sec. \ref{SubSec:Injection_model}. In this framework we find $\eta _{A399} =  0.80 \pm 0.05$ and $\eta _{A401} = 0.66 \pm 0.04 $. The global scaling indices of the magnetic fields we find in the re-acceleration scenario are slightly higher than the ones we found with the injection model. The scaling indices of the magnetic fields in the two scenarios are summarized in Table \ref{tab:eta}.

\begin{table}
    \caption{Overview of the best fit parameters for the scaling index of the magnetic field in the two different scenarios. Uncertainties inside the brackets reflect the dependence of $\eta$ with respect to the parameters of the $\beta$-profile; see Sec. \ref{SubSec:Injection_model}.}
    \centering
    \begin{center}
    \begin{tabular}{cccc}
        
        \multicolumn{4}{c}{ \textbf{radio--X--SZ} } \\
        \hline
        \hline
        Region & Injection Model & Re-Acceleration Model & $\rho$\\
        \hline
        A401 & $0.58 \pm 0.03 \left( \pm 0.03 \right)$ & $0.66 \pm 0.04 \left( \pm 0.03 \right) $ & 0.90 \\
        
        A399 & $0.70 \pm 0.04 \left( \pm 0.01 \right)$ & $0.80 \pm 0.05 \left( \pm 0.01 \right)$ & 0.74\\
        \hline
        \hline
        
    \end{tabular}
    \end{center}

    \label{tab:eta}
\end{table}

\subsection{Estimation of $\eta _{A401}$ from rotation measure}
\label{sec:MagneticField_RM}

\begin{figure*}
    \centering
    \includegraphics[scale=0.54]{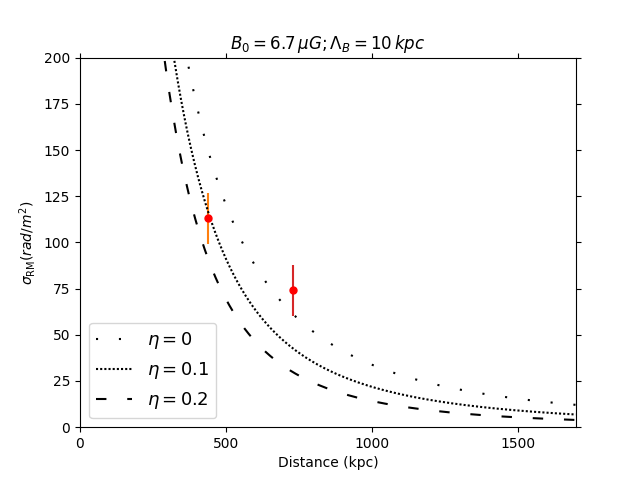}
    \includegraphics[scale=0.54]{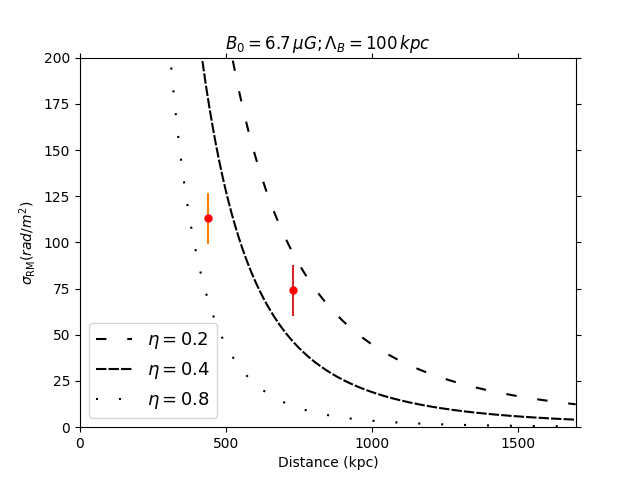}
    \includegraphics[scale=0.54]{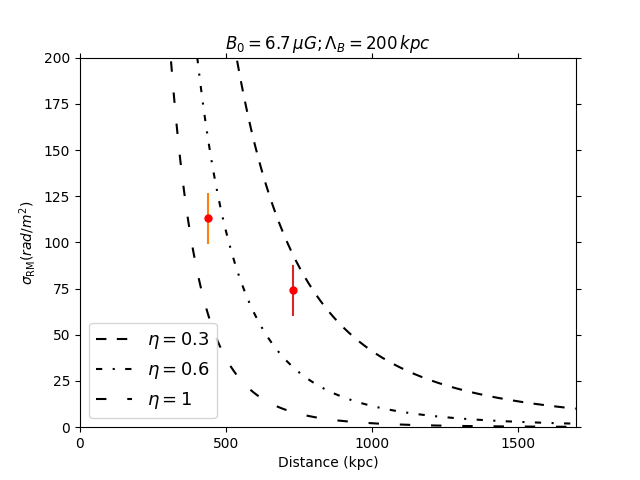}
    \includegraphics[scale=0.54]{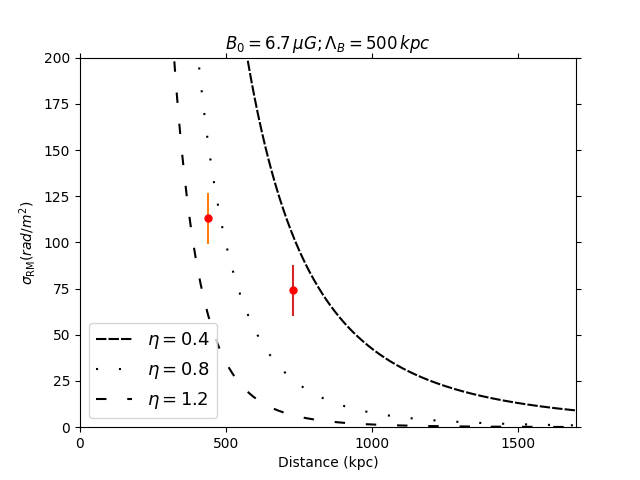}
    \caption{RMS of the RM profile as a function of the distance from the center of A401 for three different values of the magnetic field scaling index $\eta$. Red points indicate the RM values in A401 \citep{Govoni10}. \textbf{Top Left:} $\Lambda _B = 10$\,kpc. \textbf{Top Right:} $\Lambda _B = 100$ kpc. \textbf{Bottom Left:} $\Lambda _B = 200$\,kpc. \textbf{Bottom Right:} $\Lambda _B = 500$\,kpc.}
    \label{fig:eta_401_Multi}
\end{figure*}

\begin{figure}
    \centering
    \includegraphics[scale=0.5]{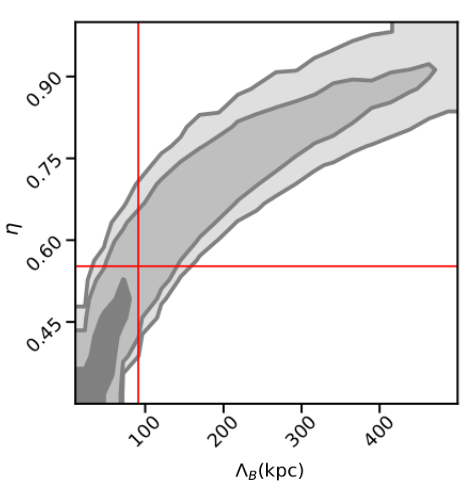}
    \caption{The fit of the RMS of the RM values for two radio galaxies in A401, red lines indicate the best fit values corresponding to the $50^{\mathrm{th}}$ percentile. Grey contours from lightest to darkest refer to the 3, 2, and 1$\sigma$ levels in the 2D histogram.}
    \label{fig:eta_vs_LambdaB}
\end{figure}

We can estimate $\eta$ with an independent, but approximate, method using ancillary rotation measure (RM) data on A401 from \cite{Govoni10}.
Considering a density profile that follows a $\beta$--model, the RM dispersion profile  can be written as \citep[and references therein]{dolag01}:
\begin{equation}
    \sigma _{\mathrm{RM}} (r) = \frac{K B_{0} n_{0} r_{\mathrm{c}}^{1/2} \Lambda _B ^{1/2}}{\left( 1+r^2/r _c ^2 \right)^{(6\beta(1+\eta)-1)/4}} \sqrt{\frac{\Gamma(3\beta(1+\eta) - 1/2)}{\Gamma(3\beta(1+\eta))}},
    \label{eq:sigma_rm(r)}
\end{equation}
where $B_{0}$ is the magnetic field at the cluster center, $\Gamma$ is the gamma function, $n_0$ is the central density, $\Lambda_{B}$ is the magnetic field auto-correlation length, and \textit{K} is a dimensionless constant that depends on the integration path through the gas of the cluster. If we assume that the source is located near the middle of the cluster, the value of the constant is $K=441$, while if the source is entirely beyond the cluster, $K=624$ \citep{dolag01}. In this work, we assume the former case.

By making some assumptions on these parameters, we plot $\sigma _{\mathrm{RM}}$ versus radius for different values of $\eta$ and $\Lambda_B$ and compare them with the RM values measured by \cite{Govoni10}. In this way, we can determine $\eta$.
First, we estimate the magnetic field strength at the center of A401 using the relation discovered by \cite{Govoni17} between the central magnetic field and the gas density: $B_{0} \simeq C \left( \frac{n_{0}}{\rm{cm}^{-3}} \right) ^{0.47}$, where $C = 69.5 \, \mu$G. Note that this relationship statistically links the magnetic field central strength with the thermal gas density at the center of the cluster but does not give any information about the magnetic field radial decline. By using the central gas density derived by \cite{hincks22} for A401 as best fit parameter, we estimate $B_{0,\,A401} \sim 6.7 \, \mu$G. Note that using the central gas density in the filament estimated by the previous authors, we derive $B_{\mathrm{fil}} \sim 0.8 \, \mu$G, in agreement with the upper limit established by \cite{Govoni19}. However, this estimation must be interpreted with caution because the relation used has been derived from the denser environment in clusters that may not be accurate in the case of filaments.
Finally, for the magnetic field auto-correlation length \citep{ensslin2003,murgia2004}:
\begin{equation}
    \Lambda _B = \frac{3 \pi}{2} \frac{\int _{k_{\rm min}} ^{k_{\rm max}} |B_{k}|^{2} k dk}{\int _{k_{\rm min}} ^{k_{\rm max}} |B_{k}|^{2} k^2 dk},
 \end{equation}  
where $k _{\rm min}$ and $k_{\rm max}$ are the minimum and maximum wave-numbers ($k=2\pi/\Lambda$ where $\Lambda$ indicates the scale of fluctuation) of the magnetic field and $|B_{k}|^{2}$ is the magnetic field power spectrum. We consider three cases: $\Lambda _{B} = $ 100, 200 and 500\,kpc. Indeed, typical values of the magnetic field auto-correlation length in merging galaxy clusters can be on the order of a few hundred kpc \citep[e.g., ][]{Vacca10,girardi2016}. Moreover, to exclude the possibility that $\Lambda_B$ is only a few kpc, we also test the case $\Lambda_B = 10$\,kpc. Using equation~(\ref{eq:sigma_rm(r)}) we extract the $\sigma _{\mathrm{RM}}$ profile for different values of the magnetic field scaling index and auto-correlation scale that can be compared with the RM values measured by \cite{Govoni10} for two radio galaxies in A401. The RM dispersion profiles and the measures are shown in Fig. \ref{fig:eta_401_Multi}. These plots indicate that values of $\eta$ spanning between $\sim$0.3--1.0 provide a good description of the rotation measure observed in the cluster, while smaller and larger values appear to be excluded by the data. Consequently, the magnetic field auto-correlation length must be $\gtrsim 100$\,kpc under our assumptions. The case where $\Lambda_B = 10$ kpc is unlikely because to reproduce the observed $\sigma _{\mathrm{RM}}$ values, the magnetic field must be almost constant with radius, unlike what we know so far from simulations \citep[e.g.,][]{Dolag2002, vazza18} and observational \citep[e.g.,][]{dolag01, Murgia09} constraints. To test the previous hypothesis we fit the measures of $\sigma _{\mathrm{RM}}$ in A401 with equation~(\ref{eq:sigma_rm(r)}) as a function of $\eta$ and $\Lambda _{B}$. We assume uniform priors for the parameters, 0.3--1 for the scaling index of the magnetic field and 5--500\,kpc for the auto-correlation scale. Computing the 16$^{\mathrm{th}}$, 50$^{\mathrm{th}}$, and 84$^{\mathrm{th}}$ percentiles of the samples in the marginalized distributions, we find $\Lambda _{B} = 83 ^{+164} _{-45}$ kpc and $\eta = 0.53 ^{+0.25} _{-0.17}$.  Despite the significant uncertainties, due to the low number of data and degeneracy between the two parameters, these values agree with the scaling index of the magnetic field we found in Sec.~\ref{SubSec:Injection_model} and \ref{SubSec:Re-acc}. However, we can not make any statement about the physical mechanism responsible for the non-thermal electron population. We show in Fig. \ref{fig:eta_vs_LambdaB} the posterior distribution of the two parameters. Considering this plot together with the paucity of available RM data, we conclude that previous estimates of the two parameters can only be taken as an upper limit. It is clear from our study that having a larger set of RM data at several distances from the center of the system will be of fundamental importance to reduce the degeneracy between the parameters and to better constrain the radial profile of $\sigma_{\rm RM} (r)$. In this context, radio interferometers such as the Jansky Very Large Array (JVLA), MeerKAT, and the Square Kilometre Array (SKA) will play a key role in mapping the magnetic field in an ever larger number of clusters. The polarisation survey planned with the SKA in band 2 (950--1760\,MHz) is expected to detect 60--90 polarised sources per square degree \citep{heald2020} and deep SKA observations in the same frequency range will uncover polarised emission from radio halos up to the cluster periphery. The simultaneous observation of polarimetric properties of discrete background radio sources as well as diffuse cluster emission will enable a detailed characterisation of the magnetic field structure, central strength and radial decrease though the cluster volume with a great degree of accuracy \citep{loi2019}. An accurate estimation of the scaling index of the magnetic field from RM data, combined with high-resolution multi-frequency observations, could finally explain the origin of non-thermal electrons within galaxy clusters.

The large radio--X peak offset we measure in the two maps for A401, ${>}100$\,kpc, seems to confirm that the magnetic field auto-correlation length is likely hundreds of kpc. Note that the radio--X peak offset can be inferred from the middle panel in Fig. \ref{fig:Maps} where we overlay the SZ contours, a good indication for the X-ray emission, on the radio map. A large auto-correlation length of the magnetic field would not be surprising. For Abell 523, a galaxy cluster with radio halo characterized by one of the largest radio--X-ray peak offsets, \cite{girardi2016} estimated that $\Lambda _{B}$ is around 300 kpc. Future multi-wavelength high-resolution observations, combined with RM data, will play a fundamental role in understanding magnetic field properties inside radio halos better.

\section{Conclusions}
\label{Sect:Conclusions}

In this paper, we have studied the point-by-point correlations in a pair of galaxy clusters, Abell 399 and Abell 401, using high-resolution radio, SZ, and X-ray maps. The radio map at 140 MHz comes from LOFAR observations, the SZ map combines ACT and $Planck$ data, while the X-ray map is a mosaic of four \XMM pointings of the A399--A401 field. Using the radio and X-ray maps, we found a sublinear relation for A401 and A399. The SZ--X comparison is in line with the profile expected for isothermal clusters, although we find an indication for weak temperature gradients in A399. Finally, using the radio and SZ maps, we discovered for the first time with ${\sim}$arcminute angular resolution a local correlation in galaxy clusters. We found that the comparison between SZ and X-ray data is much tighter and more linearly correlated than the other two. Using the radio--X and radio--SZ correlations and assuming an isothermal profile, we derived the radial link between thermal and non-thermal components in the ICM on sub-cluster scales for A401.

We combined radio, SZ, and X-ray brightnesses for the first time to derive the scaling index of the magnetic field in the two clusters, assuming a spherical $\beta$--profile for two different scenarios, the injection model and the re-acceleration model. We found that the scaling index of the magnetic field for the two clusters ranges between $\eta \sim 0.6$ and $\eta \sim 0.8$. Finally, we estimated the radial dependence of the magnetic field in A401 with an approximate method, using ancillary RM measures. We estimated $\eta \sim 0.6$ and $\Lambda_{B} \gtrsim 100$ kpc for A401. However, with the current data sets we cannot make any claims about the physical mechanism responsible for the non-thermal electron population inside clusters.

The large sample of galaxy clusters resolved both by LOFAR, MeerKAT, and ACT will allow us to further investigate these correlations for a larger population of systems. Including X-ray data and RM measures, we aim to increase our knowledge of the physical mechanisms responsible for the non-thermal electron population in radio halos and to constrain $\eta$ in a large sample of radio halos.

\section*{Acknowledgements}

We thank the anonymous referee whose comments helped to improve
the quality of the paper. We thank Paolo de Bernardis and Silvia Masi for useful suggestions which improved the quality of the article.

The ACT project is supported by the U.S. National Science Foundation through awards AST-0408698, AST-0965625, and AST1440226, as well as awards PHY-0355328, PHY-0855887 and PHY1214379. Funding was also provided by Princeton University, the University of Pennsylvania, and a Canada Foundation for Innovation (CFI) award to UBC. ACT operates in the Parque Astronómico Atacama in northern Chile under the auspices of the La Agencia Nacional de Investigación y Desarrollo (ANID; formerly Comisión Nacional de Investigación Científica y Tecnológica de Chile, or CONICYT). The development of multichroic detectors and lenses was supported by NASA grants NNX13AE56G and NNX14AB58G. Detector research at NIST was supported by the NIST Innovations in Measurement Science program. Computations were performed on Cori at NERSC as part of the CMB Community allocation, on the Niagara supercomputer at the SciNet HPC Consortium, and on Feynman and Tiger at Princeton Research Computing, and on the hippo cluster at the University of KwaZulu-Natal. SciNet is funded by the CFI under the auspices of Compute Canada, the Government of Ontario, the Ontario Research Fund–Research Excellence, and the University of Toronto. Colleagues at AstroNorte and RadioSky provide logistical support and keep operations in Chile running smoothly. We also thank the Mishrahi Fund and the Wilkinson Fund for their generous support of the project.

This paper is based on data obtained with the International LOFAR Telescope (ILT) under project code LC2\_005. LOFAR \citep{Lofar2013} is the Low Frequency Array designed and constructed by ASTRON. It has observing, data processing, and data storage facilities in several countries, that are owned by various parties (each with their own funding sources), and that are collectively operated by the ILT foundation under a joint scientific policy. The ILT resources have benefit.

This work is based on observations obtained with \XMM, an European Space Agency (ESA) science mission with instruments and contributions directly funded by ESA Member States and NASA.\\
VV and MM acknowledge support from INAF mainstream project “Galaxy Clusters Science with LOFAR” 1.05.01.86.05. FV acknowledges financial support by the ERC Starting Grant “MAGCOW" (n. 714196). FL acknowledges financial support from the Italian Ministry of University and Research - Project Proposal CIR01\_00010. KM and MJH acknowledge support from the National Research Foundation of South Africa. AB acknowledges support from the ERC-Stg program DRANOEL 714245. ADH acknowledges support from the Sutton Family Chair in Science, Christianity and Cultures and from the Faculty of Arts and Science, University of Toronto.

Some of the results in this paper have been obtained using the following packages: 
\textsc{Astropy},\footnote{http://www.astropy.org} a community-developed core Python package for Astronomy \citep{software:astropy/a,software:astropy/b};
\textsc{ds9} \citep{software:ds9};
\textsc{emcee} \citep{foreman13};
\textsc{Matplotlib} \citep{software:matplotlib};
\textsc{NumPy} \citep{software:numpy}
and \textsc{SciPy} \citep{software:scipy}.
This research made use of \textsc{Montage}. It is funded by the National Science Foundation under Grant Number ACI-1440620, and was previously funded by the National Aeronautics and Space Administration's Earth Science Technology Office, Computation Technologies Project, under Cooperative Agreement Number NCC5-626 between NASA and the California Institute of Technology.

\section*{Data Availability}

The ACT+\Planck $y$-map used in this paper will be released on the NASA Legacy Archive Microwave Background Data Analysis (LAMBDA) website. The \XMM map used in this work has been obtained analyzing publicly available data; however, the final product will be shared on reasonable request to the corresponding author. The LOFAR map will be released on reasonable request to F. Govoni.

\bibliographystyle{mnras}
\bibliography{bridge_correlation_ACT.bib}

\appendix
\section{Further considerations on the radio--x and radio--SZ correlations}
\label{appendix:masking_clusters}

\begin{figure*}
    \centering
    {\includegraphics[scale=0.6]{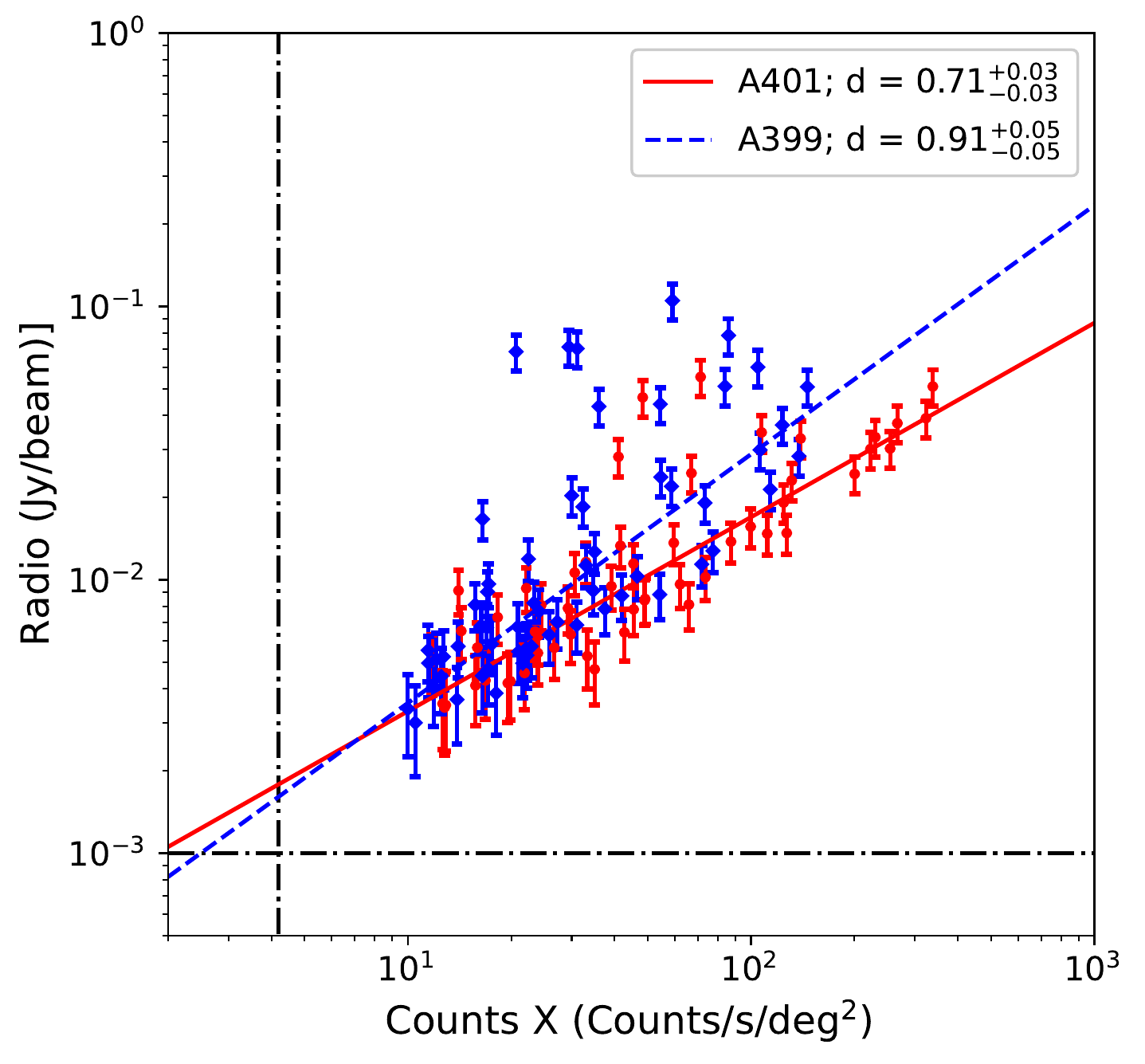}}
    {\includegraphics[scale=0.6]{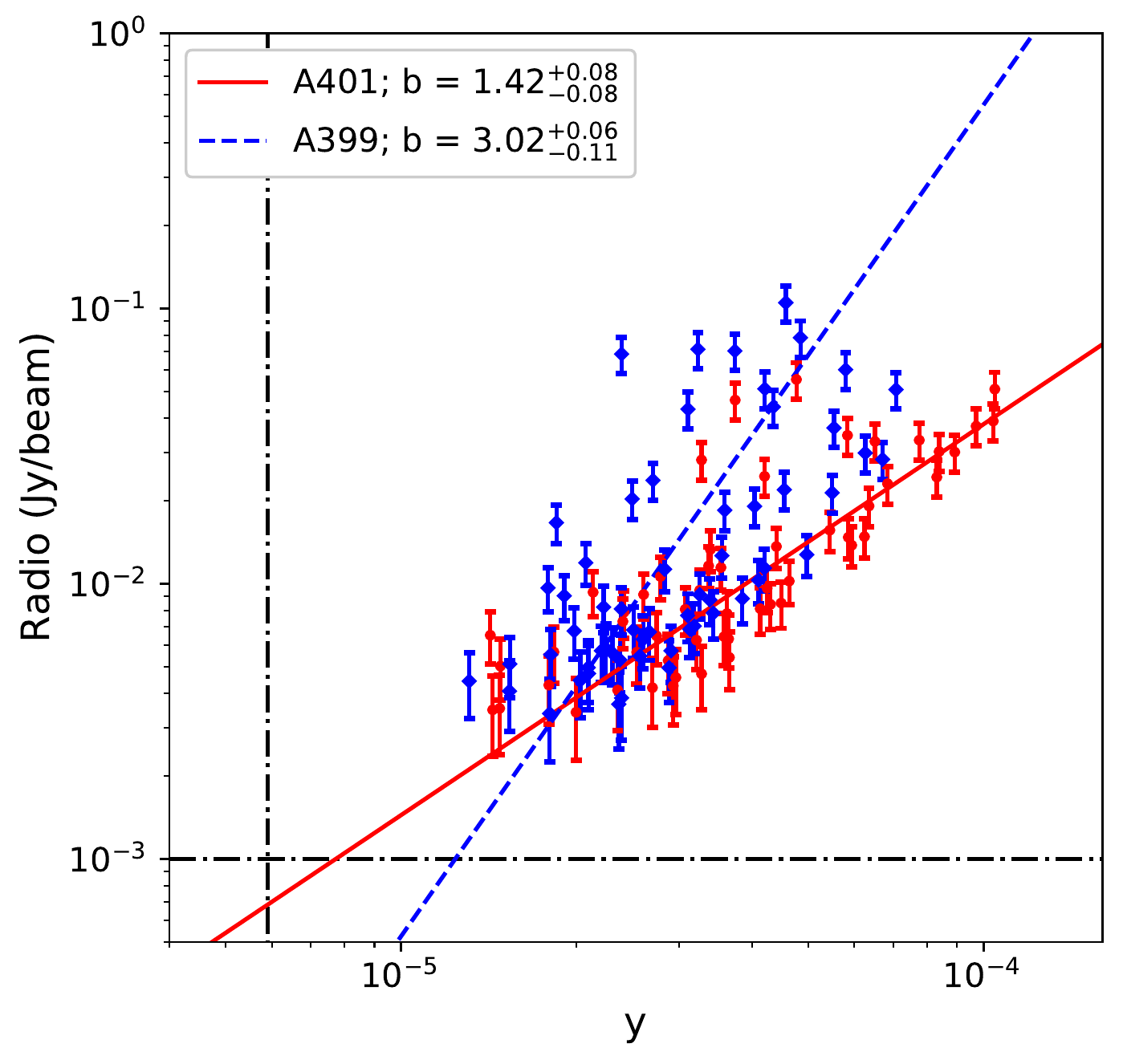}}
    \caption{\textbf{Left}: Correlation between the radio and X-ray count rate maps. \textbf{Right}: Correlation between the radio map and the $y$ signal. The red points and the solid line refer to A401, while the blue diamond markers and the dashed line refer to A399. With respect to Fig. \ref{fig:Correlation_SZ_radio}, we mask the up-scattered radio points in A401 and the east sector in A399.
    In the three figures, we show only the points that satisfy the condition $S/N > 2$. The horizontal and vertical dash-dotted lines indicate the noise level in the two maps.}
    \label{fig:Correlation_SZ_radio_Masked}
\end{figure*}

We present the radio--X and radio--SZ correlations for A399 and A401 after masking both the east sector of A399, which has a sharp radio edge emission, and the possibly up-scattered radio points in A401. In Sec. \ref{SubSec:X-radio}, after having analysed high spatial resolution radio and optical maps, we did not find any compact source at the spatial location of the up-scattered radio points in A401. Despite this, we can not have absolute certainty that the up-scattered radio points belong to the diffuse emission associated with the ICM; for this reason, here we investigate the impact of these points on the global fit. The resulting correlations, using a threshold of $S/N>2$, are shown in Fig.~\ref{fig:Correlation_SZ_radio_Masked}. Fit results for $S/N>2$ threshold are summarized in Table~\ref{tab:fit_params_mask}.

In A399, we find that when we mask the east sector, the scaling indices for all the correlations are at least ${\sim}2\sigma$ compatible with the ones we extract in the whole cluster. Moreover, we find that the Pearson correlation coefficients for the radio--X and radio--SZ relations are closer to unity. As expected, masking the east sector of A399 reduces the scatter of the data and results in a much more linearly correlated trend in the double log scale. For A401, masking the up-scattered radio points yields radio--X and radio--SZ correlations that are more linearly correlated and $1 \sigma$ compatible with respect to those reported in Sec.~\ref{Sect:Correlation}. Masking the two regions result in a much tighter correlation but does not drastically affect the results we have presented in the main text.

\begin{table*}
    \caption{Overview of the best-fit parameters using the $S/N>2$ cut-off for the radio--SZ, radio--X and SZ--X correlations in A401, A399, and the filament masking the east sector of A399 and the up-scattered radio data in A401.}
    \centering
    \begin{center}
    \begin{tabular}{cccc|ccc|ccc}
        \multicolumn{10}{c}{ \textbf{$S/N>2$} } \\ 
         & \multicolumn{3}{c}{ \textbf{radio--SZ} $\left( \mathrm{equation}~\ref{eq:radio-SZ_corr} \right)$ } & \multicolumn{3}{c}{ \textbf{radio--X} $\left( \mathrm{equation}~\ref{eq:radio-x_corr} \right)$} & \multicolumn{3}{c}{ \textbf{SZ--X} $\left( \mathrm{equation}~\ref{eq:SZ-X_corr} \right)$ } \\
        \hline
        \hline
        \multicolumn{1}{c|}{Region} & a$_{\rm radio-SZ}$ & b & $\rho$ & a$_{\rm radio-X}$ & d & $\rho$ & a$_{\rm SZ-X}$ & c & $\rho$\\
        \hline
        \multicolumn{1}{c|}{A401} & $ 4.26 ^{+0.49} _{-0.23} $ & $1.42 ^{+0.08} _{-0.08}$ & 0.80 & $ -3.18 ^{+0.06} _{-0.06}$ & $0.71 ^{+0.03} _{-0.03}$ & 0.84 
        & $ -5.27 ^{+0.04} _{-0.04}$ & $0.51 ^{+0.02} _{-0.02}$  & 0.97\\
        
        \multicolumn{1}{c|}{A399} & $ 11.82 ^{+0.34} _{-0.30} $ & $3.02 ^{+0.06} _{-0.11}$ & 0.68 & $ -3.35 ^{+0.09} _{-0.08}$ & $0.91 ^{+0.05} _{-0.05}$ & 0.72 & 
        $ -5.29 ^{+0.05} _{-0.05}$ & $0.51 ^{+0.03} _{-0.03}$  & 0.94\\
        
        \hline
    \end{tabular}
    \end{center}
    
    \label{tab:fit_params_mask}
\end{table*}

\section{Correlations in double log scale}
\label{appendix:log_fit}

In Sec. \ref{Sect:Correlation} we fit the correlation in a linear scale with a power law over the whole clusters. Here we analyze again the radio--X, radio--SZ, and SZ--X correlations fitting the data in the log--log plane with a straight line to ensure our results conform to the usual practice in the literature. The best fit parameters using a $S/N>2$ cut-off are summarized in Table~\ref{tab:fit_params_log}. We note the agreement between the SZ--X correlations in the two clusters, indicating that the two methods are almost equivalent when the scatter of the data is negligible. Conversely, the main differences between the two methods reside in the radio--SZ correlation for A401 ($>5 \sigma$ tension) and the radio--SZ and radio--X correlations for A399 ($<3 \sigma$ tension).

In the case of the radio--SZ correlation for A401, fitting the data in the double log scale, we derive a much steeper relation due to the up-scattered radio points. When we mask these regions, we find that the best-fit parameters fully agree with the results we derive in Sec. \ref{SubSec:SZ-radio}. The radio--SZ and radio--X correlations in A399 we report here are flatter than the ones we extract fitting the data in the linear space. We conclude that deriving the best-fit parameters in the log--log plane only works well when the correlations are tight and not scattered. Physically our test indicates that the most straightforward linear fit in the double log plane should be avoided in the presence of irregular clusters, residual compact sources, and low signal-to-noise observations.

\begin{table*}
    \caption{Overview of the best-fit parameters using $S/N>2$ cut-off for the radio--SZ, radio--X, and SZ--X correlations in A401, A399 fitting the data in a double log scale with a straight line.}
    \centering
    \begin{center}
    \begin{tabular}{ccc|cc|cc}
        \\
        \multicolumn{7}{c}{ \textbf{$S/N>2$} } \\
         & \multicolumn{2}{c}{ \textbf{radio--SZ} } & \multicolumn{2}{c}{ \textbf{radio--X} } & \multicolumn{2}{c}{ \textbf{SZ--X} } \\
        \hline
        \hline
        \multicolumn{1}{c|}{Region} & a$_{\rm radio-SZ}$ & b & a$_{\rm radio-X}$ & d & a$_{\rm SZ-X}$ & c\\
        \hline
        \multicolumn{1}{c|}{A401} & $ 7.65 ^{+0.47} _{-0.44} $ & $2.19 ^{+0.11} _{-0.10}$ & 
        $ -3.17 ^{+0.05} _{-0.05}$ & $0.78 ^{+0.03} _{-0.03}$ & 
        $ -5.24 ^{+0.04} _{-0.04}$ & $0.50 ^{+0.02} _{-0.02}$\\
        
        \multicolumn{1}{c|}{A399} & $ 8.68 ^{+0.63} _{-0.58} $ & $2.34 ^{+0.14} _{-0.13}$  & 
        $ -2.70 ^{+0.04} _{-0.05}$ & $0.64 ^{+0.03} _{-0.03}$ & 
        $ -5.26 ^{+0.05} _{-0.05}$ & $0.50 ^{+0.03} _{-0.03}$\\

        \hline
    \end{tabular}
    \end{center}
    
    \label{tab:fit_params_log}
\end{table*}

\input{affiliations}

\bsp	
\label{lastpage}
\end{document}

%% file: authors.tex
\author[F.~Radiconi,~V.~Vacca~et~al.]{%
Federico~Radiconi,$^{1,2}$ \thanks{E-mail: federico.radiconi@roma1.infn.it}
Valentina~Vacca,$^{3}$ \thanks{E-mail: valentina.vacca@inaf.it}
\newauthor
Elia~Battistelli,$^{1,2}$
Annalisa~Bonafede,$^{4,5}$
Valentina~Capalbo,$^{1}$
Mark~J.~Devlin,$^{6}$
Luca~Di~Mascolo,$^{7,8,9}$
\newauthor
Luigina~Feretti,$^{5}$
Patricio~A.~Gallardo,$^{10}$
Ajay Gill,$^{11}$
Gabriele~Giovannini,$^{4,5}$
Federica~Govoni,$^{3}$
\newauthor
Yilun~Guan,$^{11}$
Matt~Hilton,$^{12,13}$
Adam~D.~Hincks,$^{14,15}$
John~P.~Hughes,$^{16}$
Marco~Iacobelli,$^{18}$
\newauthor
Giovanni~Isopi,$^{1}$
Francesca~Loi,$^{3}$
Kavilan Moodley,$^{12,13}$
Tony~Mroczkowski,$^{17}$
Matteo~Murgia, $^{3}$
\newauthor
Emanuela~Orr\'u,$^{18}$
Rosita~Paladino,$^{5}$
Bruce~Partridge,$^{19}$
Craig L. Sarazin,$^{20}$
Jack Orlowski Scherer,$^{6}$
\newauthor
Crist\'obal Sif\'on,$^{21}$
Cristian Vargas,$^{22,23}$
Franco Vazza$^{4,5}$~and
Edward~J.~Wollack$^{24}$
\newauthor
\newline
\textit{\normalsize Affiliations are listed at the end of the paper.}%
}%

%% file: affiliations.tex
\section*{Affiliations}
\noindent\textit{%
$^{1}$Sapienza University of Rome,Physics Department, Piazzale Aldo Moro 5, 00185 Rome, Italy\\
$^{2}$INFN - Sezione di Roma, Piazzale Aldo Moro 5 - I-00185, Rome, Italy\\
$^{3}$INAF – Osservatorio Astronomico di Cagliari, Via della Scienza 5, I-09047 Selargius (CA), Italy\\
$^{4}$Dipartimento di Fisica e Astronomia, Università di Bologna, Via Gobetti 93/2, 40122, Bologna, Italy\\
$^{5}$INAF - Istituto di Radioastronomia, Via P. Gobetti 101, 40129 Bologna, Italy\\
$^{6}$Department of Physics and Astronomy, University of Pennsylvania, 209 South 33rd Street, Philadelphia, PA 19104, USA\\
$^{7}$Astronomy Unit, Department of Physics, University of Trieste, via Tiepolo 11, 34131 Trieste, Italy\\
$^{8}$INAF - Osservatorio Astronomico di Trieste, via Tiepolo 11, 34131 Trieste, Italy\\
$^{9}$IFPU - Institute for Fundamental Physics of the Universe, Via Beirut 2, 34014 Trieste, Italy\\
$^{10}$Kavli Institute for Cosmological Physics, University of Chicago, Chicago, IL 60637, USA\\
$^{11}$Dunlap Institute for Astronomy and Astrophysics, University of Toronto, 50 St. George St., Toronto, ON M5S 3H4, Canada\\
$^{12}$Astrophysics Research Centre, University of KwaZulu-Natal, Westville Campus, Durban 4041, South Africa\\
$^{13}$School of Mathematics, Statistics \& Computer Science, University of KwaZulu-Natal, Westville Campus, Durban4041, South Africa\\
$^{14}$David A. Dunlap Department of Astronomy \&Astrophysics, University of Toronto, 50 St. George St., Toronto ON M5S 3H4, Canada\\
$^{15}$Specola Vaticana (Vatican Observatory), V-00120 Vatican City State\\
$^{16}$Department of Physics and Astronomy, Rutgers, the State University of New Jersey, 136 Frelinghuysen Road, Piscataway, NJ 08854-8019, USA\\
$^{17}$European Southern Observatory (ESO), Karl-Schwarzschild-Strasse 2, Garching 85748, Germany\\
$^{18}$ASTRON, the Netherlands Institute for Radio Astronomy, Postbus 2, 7990 AA, Dwingeloo, The Netherlands\\
$^{19}$Department of Astronomy, Haverford College, 370 Lancaster Ave, Haverford PA 19041, USA\\
$^{20}$Department of Astronomy, University of Virginia, 530 McCormick Road, Charlottesville, VA 22904-4325, USA\\
$^{21}$Instituto de F\'isica, Pontificia Universidad Cat\'olica de Valpara\'iso, Casilla 4059, Valpara\'iso, Chile\\
$^{22}$Instituto de Astrof\'isica and Centro de Astro-Ingenier\'ia, Facultad de F\'isica, Pontificia Universidad Cat\'olica de Chile, Av. Vicu\~na Mackenna 4860, 7820436 Macul, Santiago, Chile\\
$^{23}$Department of Physics, Florida State University, Tallahassee, Florida 32306, USA\\
$^{24}$NASA / Goddard Space Flight Center, Greenbelt, MD 20771, USA
}